\documentclass[12pt]{iopart}

\usepackage{graphicx}
\usepackage{iopams}
\usepackage{wasysym}

\newcommand{\dif}{\mathrm{d}}
\newcommand{\expt}{\mathrm{e}}
\newcommand{\dfrac}{\frac}

\begin{document}

\title[Wave turbulence in preheating II]{Aspects of wave turbulence in preheating II: Rebirth of the nonminimal coupled models}

\author{J. A. Crespo and H. P. de Oliveira}

\address{Universidade do Estado do Rio de Janeiro, Instituto de F{\'{\i}}sica - Departamento de F{\'{\i}}sica Te{\'o}rica, CEP 20550-013. Rio de Janeiro, RJ, Brazil.}

\ead{jaacrespo@gmail.com, oliveira@dft.if.uerj.br}

\begin{abstract}
We study the nonlinear stage of preheating in a model consisting of a single inflaton field $\phi$ nonminimally coupled to the spacetime curvature and considering a self-coupling quartic potential $V(\phi)=\lambda \phi^4/4$. As the first motivation, we mention that this nonminimally coupled model agrees with the observational data. The second and the central issue of the present work is to exhibit aspects of wave turbulence associated with the mechanism of energy transfer from the inflaton field to the inhomogeneous fluctuations towards a state of thermalization. The imprints of the turbulence phase are mainly shown in the power spectra in time and wavenumber of relevant quantities such as the variance and the energy density of the inflaton field. We performed the simulations for several values of the nonminimally coupled models and taking into account the backreaction of the inflaton fluctuations in the dynamics of the Universe, and that allowed to determine the effective equation of state. 
\end{abstract}

%
%
\submitto{\JCAP}
%
\maketitle
%
%

\section{Introduction}

The inflationary phase is characterized by accelerated expansion of the universe in its initial stages. It is remarkable that although cosmic inflation was formulated almost 40 years ago, it has resisted to all observational tests \cite{planck18,planck13}. We highlight that the transition from an empty and cold universe to radiation dominated and hot universe constitutes a challenging enterprise to connect the early stages to the description provided by the standard cosmological model.

Preheating is an initial phase of the process of reheating after inflation characterized by a rapid transfer of energy from the inflaton field into the other matter fields \cite{basset}. The parametric resonance governs the initial stage of this enegery transfer from the inflaton field coupled to other field fluctuations and can be studied using a linear approximation. Since we can only use the linear approximation for a limited interval of time once certain field fluctuations grow exponentially, a fully nonlinear evolution must be considered.  Several works in this direction indicated the role played by the turbulence in the nonlinear stages of preheating towards the thermalization \cite{micha_03,micha_04,deol_ivano_03,dufaux_06,felder_06,crespo_14}. We point out that the turbulence here must be understood as the wave turbulence since we have no usual fluids but only fields.

The first simulations of the nonlinear phase of preheating considered the simplest models as the single field inflation with quartic self-interaction potential $V(\phi)=\lambda \phi^4/4$ \cite{kleb_97}, or the inflaton with potential $V(\phi)=m^2 \phi^2/2$ but coupled with another field \cite{prokopec}. However, according to the last survey of observational data, the oldest single field chaotic inflationary models are ruled out in particular due to the high level of the tensor-to-scalar ratio these models predict \cite{planck18}. Remarkably, the model  $V(\phi)=\lambda \phi^4/4$ including the nonminimal coupling with gravity described by $\xi \phi^2 R/2$, where $R$ is the scalar of curvature, is in perfect agreement with the observational just assuming a small absolute value of the coupling constant $\xi$ \cite{planck18,okada,kaloshi}.

In this paper, we studied the dynamics of a scalar field nonminimally coupled with gravity mentioned above taking into consideration distinct values of the coupling parameter within a range dictated by the observational data. In this case, after a successful inflationary phase, we start the evolution of the inflaton field together with its fluctuations, or inhomogeneous modes. We mention that Tsujikawa et al. \cite{tsujikawa} first studied the nonlinear preheating stage in this model taking into account the Hartree approximation in the field equation. Here, we have followed a distinct numerical approach including the backreaction inflaton fluctuations into the metric that allows us to determine the evolution of the effective equation of state. Another important aspect is that we focused on exhibiting the features of the wave turbulence through the power spectra of relevant quantities whose power-law behavior in specific regions signalizes the energy cascades from the inflaton to the inhomogeneous fluctuations. By assuming the form of the Planck distribution for the power spectrum of the inflaton energy density, we estimate the temperature at the thermalization phase.

The present work is organized as follows. In Sections I and II we present the field equations of the model and the numerical scheme, respectively. In Section IV, we explore the relevant aspects of the dynamics and the development of turbulence by presenting the power spectra of the variance and the energy density of the scalar field.  We show in Section V the reheating temperatures for several values of the coupling parameter, and in the last Section, we conclude.  

\section{The model}%

Let us consider a scalar field $\phi(\mathbf{x},t)$ nonminimally coupled with the space-time geometry and represented by the Lagrangian  \cite{tsujikawa}

\begin{equation}
\mathcal{L} = \sqrt{-g}\left[\dfrac{R}{2m_{pl}^2} - \dfrac{1}{2} \nabla_\mu \phi \nabla^\mu \phi - \dfrac{1}{2}\xi R \phi^2 - V\left(\phi\right)\right],\label{langrangeana_f}
\end{equation}

\noindent where $R$ is the scalar curvature, $\sqrt{-g}$ is the determinant of the space-time metric, $m_{pl}$ is the reduced Planck mass and $\xi$ is the coupling constant. The field equations reads

\begin{equation}
\fl (1-\xi m_{pl}^{-2} \phi^2)G_{\mu\nu} = \frac{1}{m_{pl}^2}\left\lbrace (1-2 \xi)\partial_\mu \phi \partial_\nu \phi - g_{\mu\nu} \left[\left(\frac{1}{2}-2\xi\right)(\partial \phi)^2 - V(\phi) + 2\xi\phi \square \phi \right]\right\rbrace
\end{equation}

\begin{equation}
\square \phi - \xi R \phi - \frac{\partial V}{\partial \phi} = 0, \label{kg_eq0}
\end{equation}

\noindent where $\square \phi = g^{\mu\nu} \nabla_\mu(\partial_\nu \phi)$. We consider the following self-coupling potential

\begin{equation}
V(\phi)  = \frac{1}{4} \lambda \phi^4. \label{pot_phi4}
\end{equation}

\noindent This model can be reconciled with the observational data for $\xi \leqslant -10^{-3}$ \cite{planck13,okada,kaloshi}.

The spacetime is described by the Friedmann-Lema{\^\i}tre-Robertson-Walker (FLRW) flat spacetime with the line element

\begin{equation}
\dif s^2 = - \dif t^2 + a^2\left(t\right) \dif \mathbf{x}. \dif \mathbf{x}, \label{metrica_flrw}
\end{equation}

\noindent where $a(t)$ is the scale factor. The scalar field $\phi(\mathbf{x},t)$ has a homogeneous component, $\phi_0(t)$ that plays a relevant role during the inflationary phase, and an inhomogeneous component, $\delta \phi(\mathbf{x},t)$ due to the quantum fluctuations of the scalar field. The dymanmics of the spacetime has the contribution of the small inhomogeneites according with the following form of the field equations

\begin{equation}
G_{\mu\nu} = \frac{1}{m_{pl}^2} \left\langle T_{\mu\nu}^{\mathrm{eff}}\right\rangle, \label{einstein_eq}
\end{equation}

\noindent where $\left\langle ... \right\rangle$ denotes the spatial average in the physical domain and $T_{\mu\nu}^{\mathrm{eff}}$ is the effective energy-momentum tensor obtained after writting the Eq. (3) in the form shown above. We impose the condition $\left\langle \delta \phi(\mathbf{x},t) \right\rangle = 0$ for the fluctuations.

We introduce dimensionless variables to write the field equations for the numerical integration. For the model under consideration, we have

\begin{equation}
\mathbf{x}_p = \sqrt{\lambda}\phi_e\mathbf{x}, \quad \dif t_p = \sqrt{\lambda}\phi_e a^{-1} \dif t, \quad \phi_p =  a \phi^{-1}_e \phi. \label{adim_var_phi4}
\end{equation}

\noindent The subscript $p$ indicates the dimensionless version of the variable in the computational domain, $\phi_e=\phi_0(t_{\mathrm{end}})$ is the amplitude of the homogeneous inflaton field at the end of inflation, and $t_p \equiv \tau$ is the conformal dimensionless time. With these new definitions, the Klein-Gordon equation (\ref{kg_eq0}) becomes 

\begin{eqnarray}
\phi''_p - \dfrac{a''}{a}\phi_p - \nabla_p^2 \phi_p + \phi_p^3 +\frac{\xi a^2}{\lambda \phi_e^2}\left\langle R \right\rangle= 0, 
\end{eqnarray}

\noindent where $\phi' \equiv \partial \phi/ \partial \tau$ and $\nabla^2_p$ is the flat three-dimensional Laplacian operator. Since 

\begin{eqnarray}
\left\langle R \right\rangle = 6\left(\frac{\ddot{a}}{a}+\frac{\dot{a}^2}{a^2}\right)=6\lambda \phi_e^2 \frac{a''}{a^3},
\end{eqnarray}

\noindent it follows that

\begin{eqnarray}
\phi''_p - \left(1 - 6\xi\right)\dfrac{a''}{a}\phi_p - \nabla_p^2 \phi_p + \phi_p^3 = 0. \label{kg_phi_4}
\end{eqnarray}

\noindent The remaining relevant field equations are

\begin{eqnarray}
\fl H^2_p & \fl \qquad = & \fl \qquad \qquad \frac{\phi^2_e}{3 m_{pl}^2} \left\langle \frac{\frac{1}{2} \left( \frac{\phi_p}{a}\right)'^2+ \frac{(\boldsymbol{\nabla}_p \phi_p)^2}{a^2} + \frac{\phi^4_p}{4 a^2} + \frac{6 \xi H_p}{a^2}\left( \frac{\phi_p}{a}\right)' - \frac{2 \xi}{a^2} \nabla^2_p \phi^2_p}{1-\xi m^{-2}_{pl}a^{-2} \phi^2_e \phi^2_p} \right\rangle \label{eq_h_phi4} \\
\nonumber\\
\fl \dfrac{a''}{a} & \fl \qquad = & \fl \qquad \quad  -\dfrac{\left(1-6\xi\right)\phi_e^2}{6m_{{pl}}^2}\left\langle \dfrac{-\left( \frac{\phi_p}{a}\right)'^2  +\frac{(\boldsymbol{\nabla}_p \phi_p)^2}{a^2} + \frac{\phi^4_p}{a^2}}{1 - \xi\left(1 - 6\xi\right)\phi_e^2m_{{pl}}^{-2}a^{-2}\phi_p^2} \right\rangle, \label{eq_a_phi4} 
\end{eqnarray}

\noindent where $H_p = a'/a$. 

We integrate numerically the system of field equations (\ref{kg_phi_4}) and (\ref{eq_a_phi4}) starting at the end of inflation and using the numerical scheme described in the next Section. We point out that Tsujikawa, Maeda and Torii \cite{tsujikawa} studied the preheating in this model in the context of the Hartree approximation and assuming a further approximation in the Hamiltonian constraint (11). Using the present notation, they have taken

\begin{eqnarray}
H^2_p = \frac{\phi_e^2}{3 m_{pl}^2}\left<\frac{...}{1-\xi m^{-2}_{pl}a^{-2} \phi^2_e \phi^2_p} \right> \approx \frac{\phi_e^2}{3 m_{pl}^2}\frac{\left\langle \hspace{1.5cm} ... \hspace{1.5cm} \right\rangle}{\left(1-\xi m^{-2}_{pl}a^{-2} \phi^2_e \left<\phi^2_p\right>\right)}. 
\end{eqnarray}

\noindent The authors of Ref. \cite{tsujikawa} pointed out the expectation that the above approximation did not affect significantly the dynamics of preheating, but it was not clear under which conditions it is valid. Then, we have opted to include in the numerical code the calculation of the spatial average present in Eqs. (\ref{eq_h_phi4}) and (\ref{eq_a_phi4}) without any approximation other than the provided by the numerical procedure.

\section{Numerical approach}

We integrate the field equations (\ref{kg_phi_4}) and (\ref{eq_a_phi4}) numerically in a three-dimensional square box of comoving size $L$ with periodic boundary conditions as a suitable model of the Universe. The algorithm based on the collocation or pseudospectral method generalizes the previous two-dimensional version \cite{crespo_14} devised to study the dynamics of several minimally coupled scalar field models in the nonlinear phase of preheating. For the sake of completeness, we present briefly the main aspects of the numerical procedure.


The scalar field is approximated by a series expansion with respect to a set of basis functions identified as the Fourier functions due to the periodic boundary conditions \cite{boyd}. Then, we have

\begin{equation}
\phi_N(\mathbf{x}_p,\tau) = \sum_{k_x,k_y,k_z=-N/2}^{N/2-1}\widehat{\phi}_\mathbf{k}(\tau) \psi_\mathbf{k}(\mathbf{x}_p)\label{base_fourier}, 
\end{equation}

\noindent where $\mathbf{k}=(k_x,k_y,k_z)$ is the comoving momentum, $N$ is the truncation order, $\widehat{\phi}_\mathbf{k}(\tau)$ represents the unknown modes. The Fourier functions $\psi_\mathbf{k}(\mathbf{x}_p)$ are

\begin{equation}
\psi_\mathbf{k}(\mathbf{x}_p) =  \expt^{\frac{2\pi i}{L_p}\mathbf{k}\cdot\mathbf{x}_p},
\end{equation}

\noindent with $L_p$ being the dimensionless size. The modes are complex functions, i. e., $\widehat{\phi}_\mathbf{k}(\tau)=\alpha_\mathbf{k}(t_p)+i\beta_\mathbf{k}(\tau)$, but satisfy the relations $\widehat{\phi}_\mathbf{k}(\tau)=\widehat{\phi}^*_\mathbf{-k}(\tau)$ to produce a real scalar field in Eq. (\ref{base_fourier}).  In addition, the spectral approximation (\ref{base_fourier}) comprises naturally the homogeneous and the inhomogeneous pieces of the scalar field:

\begin{equation}
\phi_N(\mathbf{x}_p,\tau) = \alpha_\mathbf{0}(\tau) +  \sum_{\mathbf{k} \neq \mathbf{0}}\widehat{\phi}_\mathbf{k}(\tau) \psi_\mathbf{k}(\mathbf{x}_p),
\end{equation} 

\noindent where $\alpha_\mathbf{0}(\tau)$ is the homogeneous component $\phi_0$ of the inflaton field.

The next step is to substitute the approximation (\ref{base_fourier}) into the Klein-Gordon Eq. (\ref{kg_phi_4}) to form the corresponding residual equation. According to the pseudospectral method, the unknown modes are such that the residual vanishes at the collocation, or grid, points $\mathbf{x}_{p\mathbf{n}}=L_p\mathbf{n}/N$. Therefore, we arrive at

\begin{equation}
\fl \mathrm{Res}(\mathbf{x}_{p\mathbf{n}},\tau) = \phi^{\prime\prime}_{p\mathbf{n}} - (1-6\xi) \frac{a^{\prime\prime}}{a} \phi_{p\mathbf{n}}  + \frac{4\pi}{L_p^2} \sum \vert\mathbf{k}\vert^2 \widehat{\phi}_{\mathbf{k}}(\tau) \psi_{\mathbf{k}}(\mathbf{x}_{p\mathbf{n}}) + \phi_{p\mathbf{n}}^3=0. \label{sistema_edo}
\end{equation}

\noindent In the above expression, we express the Laplacian term in function of the modes; $\phi_{p\mathbf{n}}$ are the values of the scalar field at the collocation points connected with the modes through 

\begin{equation}
\phi_{p\mathbf{n}}=\phi_N(\mathbf{x}_{p\mathbf{n}},\tau)=\sum \widehat{\phi}_{\mathbf{k}}(\tau) \psi_{\mathbf{k}}(\mathbf{x}_{p\mathbf{n}}). \label{phi_expr}
\end{equation}

\noindent We use the fast Fourier transform (FFT) to invert this relation whenever necessary.

Concerning Eq. (\ref{eq_a_phi4}) we calculate at each time step the spatial average using quadrature formulae schematically indicated by

\begin{equation}
\left\langle (...)\right\rangle = \frac{1}{L_p^3}\int_{\mathcal{D}}(...) \dif^3\mathbf{x}_p \approx \frac{1}{L_p^3}\sum_{\mathbf{n}}(...)_{\mathbf{n}} w_{\mathbf{n}},
\end{equation}

\noindent where $(...)_{\mathbf{n}}$ indicates that the corresponding term is calculated at the quadrature points and $w_{\mathbf{n}}$ are the weights. For the sake of computational economy, we made the quadrature points to coincide with the collocation points. At this point, it is important to mention other codes for exploring the nonlinear phase of preheating like the LATTICEEASY \cite{latticeeasy}, the DEFROST \cite{defrost} and the PSpectre \cite{pspectre}. Although they are very efficient, we have implemented an alternative and valid pseudospectral algorithm to taken into account the backreaction of the field fluctuations besides the power spectra in the time and space domains for the selected quantities.   
 

The integration starts at the end of inflation with $\alpha_{\mathbf{0}}(0)=1$ and  $\alpha^\prime_{\mathbf{0}}(0)=0$ for the homogeneous inflaton field, $a(0)=1$ and the initial modes and their first derivatives, $\widehat{\phi}_{\mathbf{k}}(0),\widehat{\phi}^\prime_{\mathbf{k}}(0)$, have typical amplitudes of order of $10^{-4}$ \cite{kleb_97}. We employ the FFT to obtain the values $\phi_{p\mathbf{n}}(0)$ and $\phi^\prime_{p\mathbf{n}}(0)$ associated with the inhomogeneous fluctuations and their first time derivatives necessary for the equations of motion (17). Taking these initial values we are able to calculate $a^\prime(0)$ from the Friedman equation. Therefore, with these intial values we can determine $a^\prime,\phi_{p\mathbf{n}},\phi^\prime_{p\mathbf{n}}$ at the next time level (here the FFT is used to update the modes $\widehat{\phi}_{\mathbf{k}},\widehat{\phi}^\prime_{\mathbf{k}}$) and the whole process repeats resulting in the time march. We adopted the fourth order Runge-Kutta integrator using a stepsize $h=0.001$. 

\section{Wave turbulence towards the thermalization}

We investigate the influence of the parameter $\xi$ on the aspects of wave turbulence developed in the nonlinear stage of preheating. We considered distinct values of $\xi$: $\xi=0$ representing the minimal coupling whose dynamics is already known; $\xi=-0.1$ that is in good agreement with the observational data \cite{planck13,kaloshi}; and $\xi=-1$ and $-10$ as examples of strong nonminimal couplings.

In all numerical experiments, we set $N=32$ that corresponds to a three-dimensional grid with $32^3$ points. For the sake of numerical resolution, we evolved the system in a total of $2^{24}$ time steps, or a time interval $\Delta \tau = 16777.216$. Also, we choose the self-coupling constant $\lambda = 10^{-8}$ that is consistent with the observational bounds $10^{-12} \lesssim \lambda \leqslant 10^{-4}$ imposed by the WMAP measurements to the tensor-to-scalar ratio and the spectral index \cite{planck13,okada}. It represents an alleviation of the quite small value $\approx 10^{-14}$ in the case of the minimally coupled inflaton \cite{fakir}. 

In the single field models with the quartic potential (\ref{pot_phi4}), the linear phase of preheating is governed by a Lam{\'e}-like equation with a narrow parametric resonant window. The number of resonant modes depends on the size of the box $L_p$ and the wavenumber magnitude $\vert \mathbf{k} \vert$. We choose $L_p \approx 42.7$ such that the resonant modes in the linear stage have $\vert \mathbf{k} \vert^2 = 74$.

\subsection{Interaction between the modes}

To study the energy transfer through different scales and the establishment of turbulent stages of the system, we analyze the decay of the homogeneous mode  $\alpha_\mathbf{0}(\tau)$ and the variance of the scalar field $\sigma^2_\phi(\tau)$ given by

\begin{equation}
\sigma^2_\phi = \left\langle\left(\phi_p - \left\langle\phi_p\right\rangle\right)^2\right\rangle = \sum\limits_\mathbf{k}\left\vert \widehat{\phi}_{\mathbf{k}}\right\vert^2 - \alpha_\mathbf{0}^2. \label{var_phi}
\end{equation}

\noindent This quantity gives us a relevant information about the contribution of the inhomogeneous modes of the inflaton field throughout the time evolution.

\begin{figure}[h!]
\includegraphics[scale=.58]{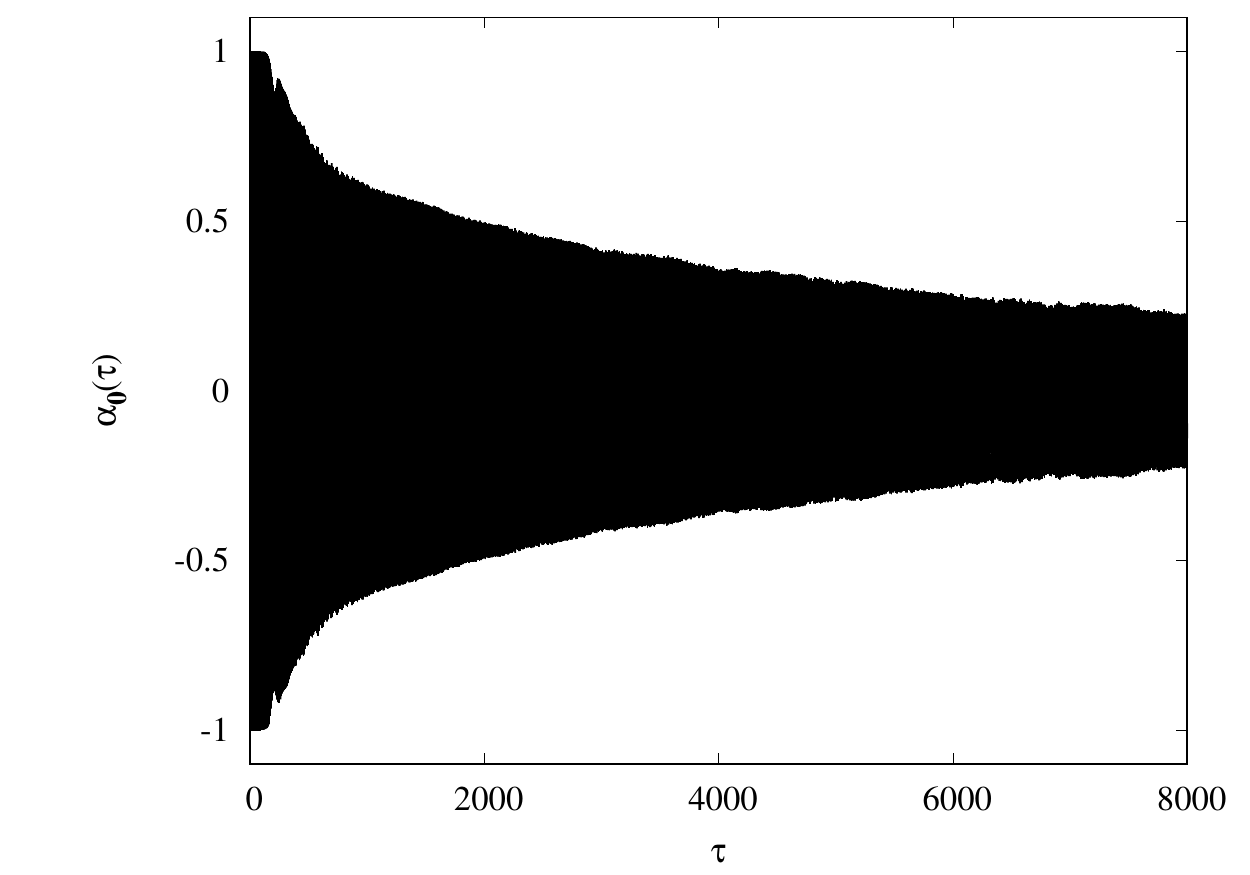}
\includegraphics[scale=.67]{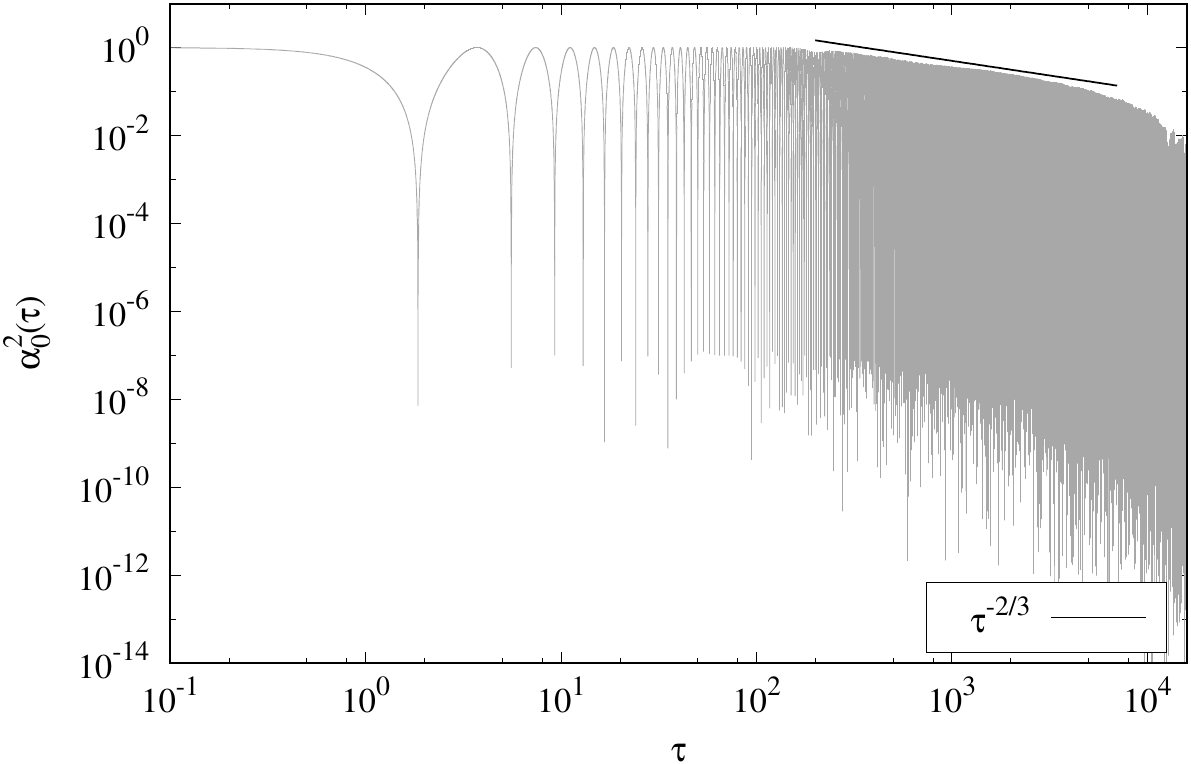}
\caption{\label{hom_0}Time evolution of the homogeneous mode, $\xi = 0$.}
\end{figure}
\begin{figure}[h!]
\includegraphics[scale=.58]{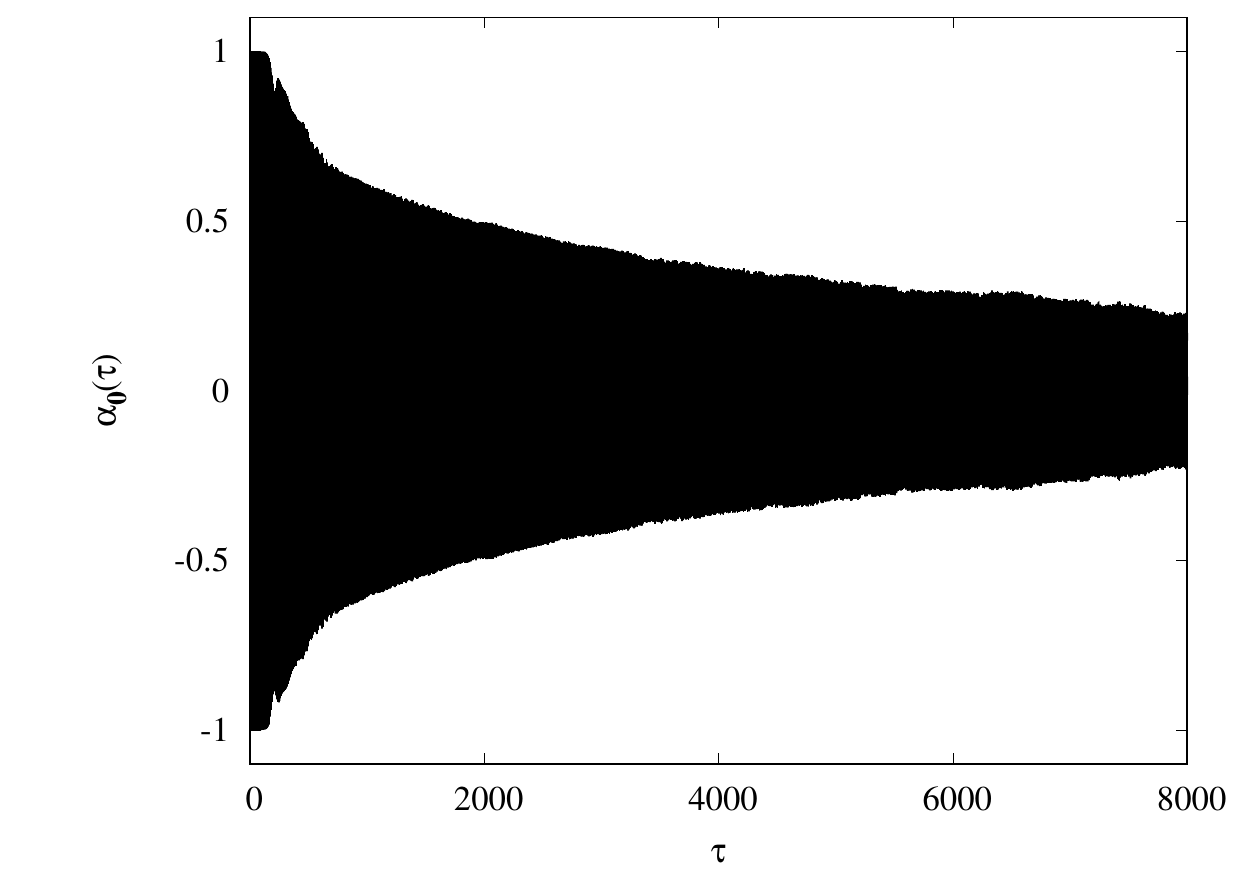}
\includegraphics[scale=.67]{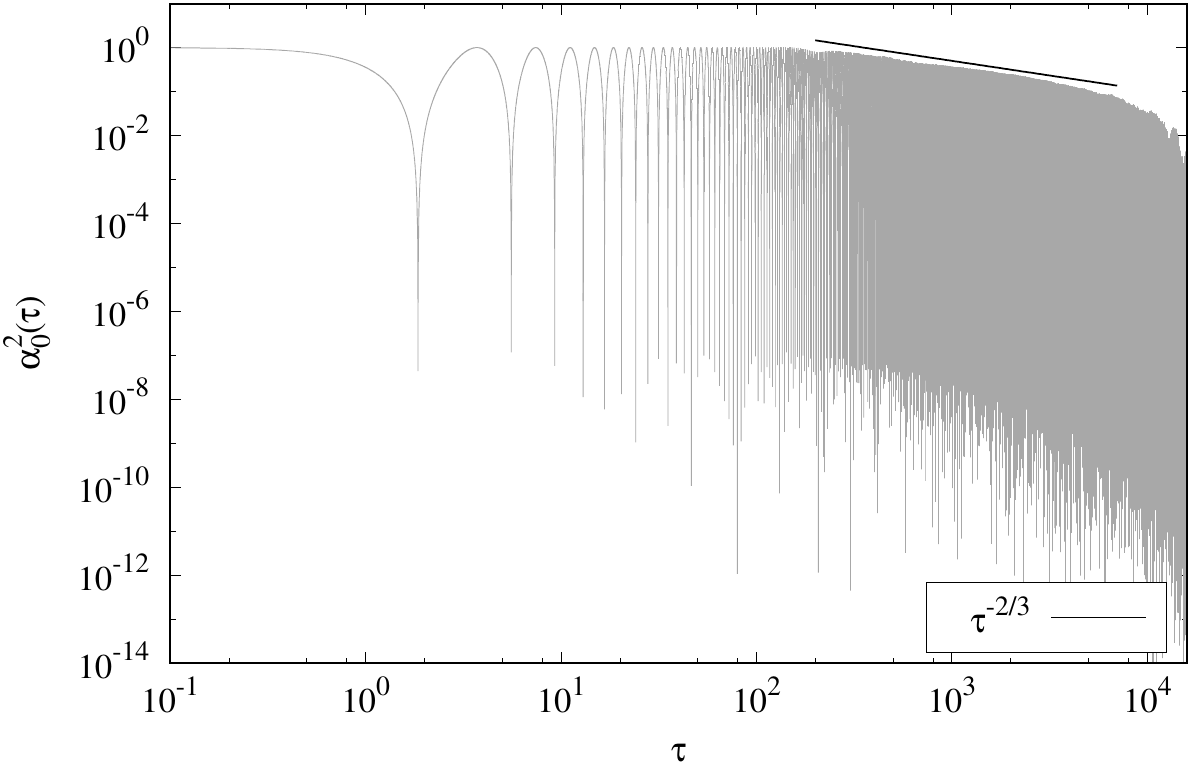}
\caption{\label{hom_01}Time evolution of the homogeneous mode, $\xi = -0.1$.}
\end{figure}
\begin{figure}[h!]
\includegraphics[scale=.58]{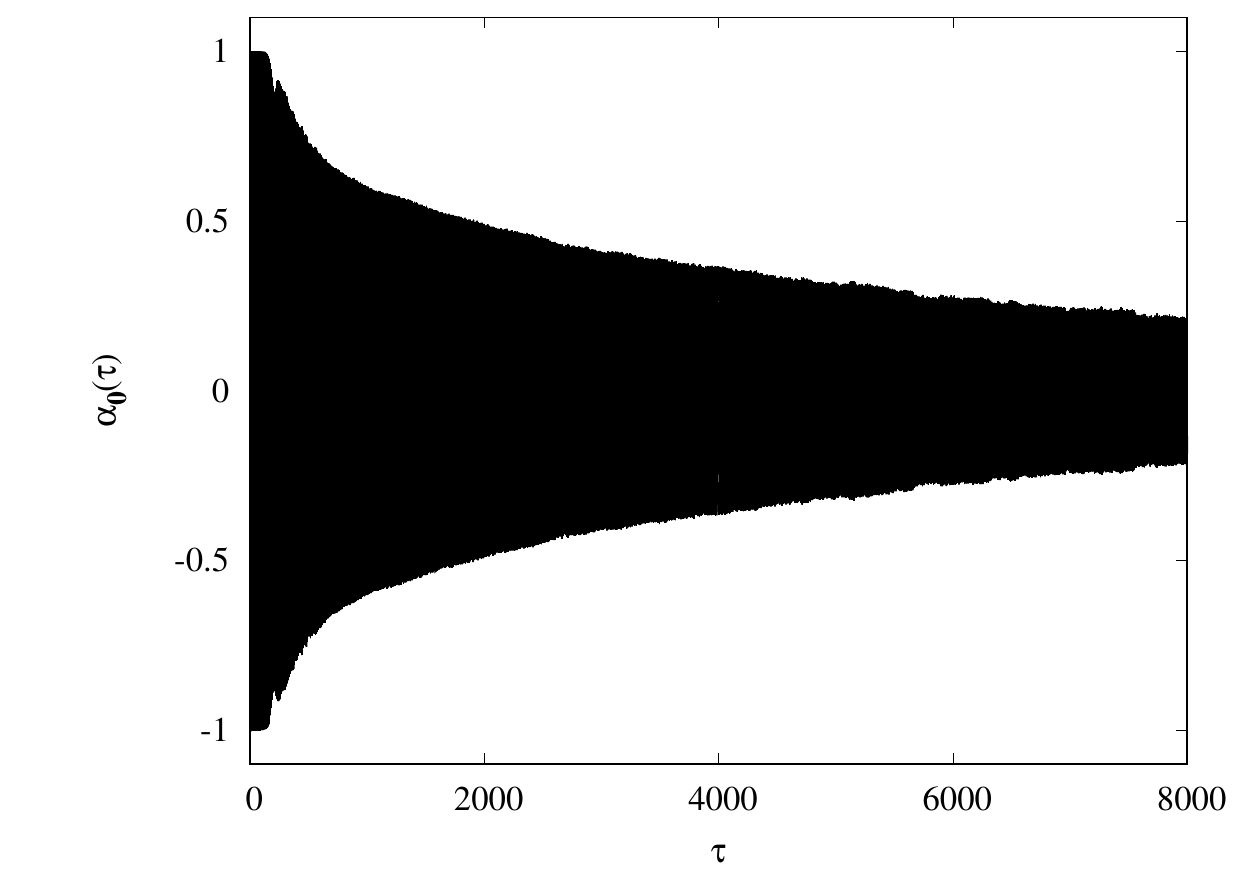}
\includegraphics[scale=.67]{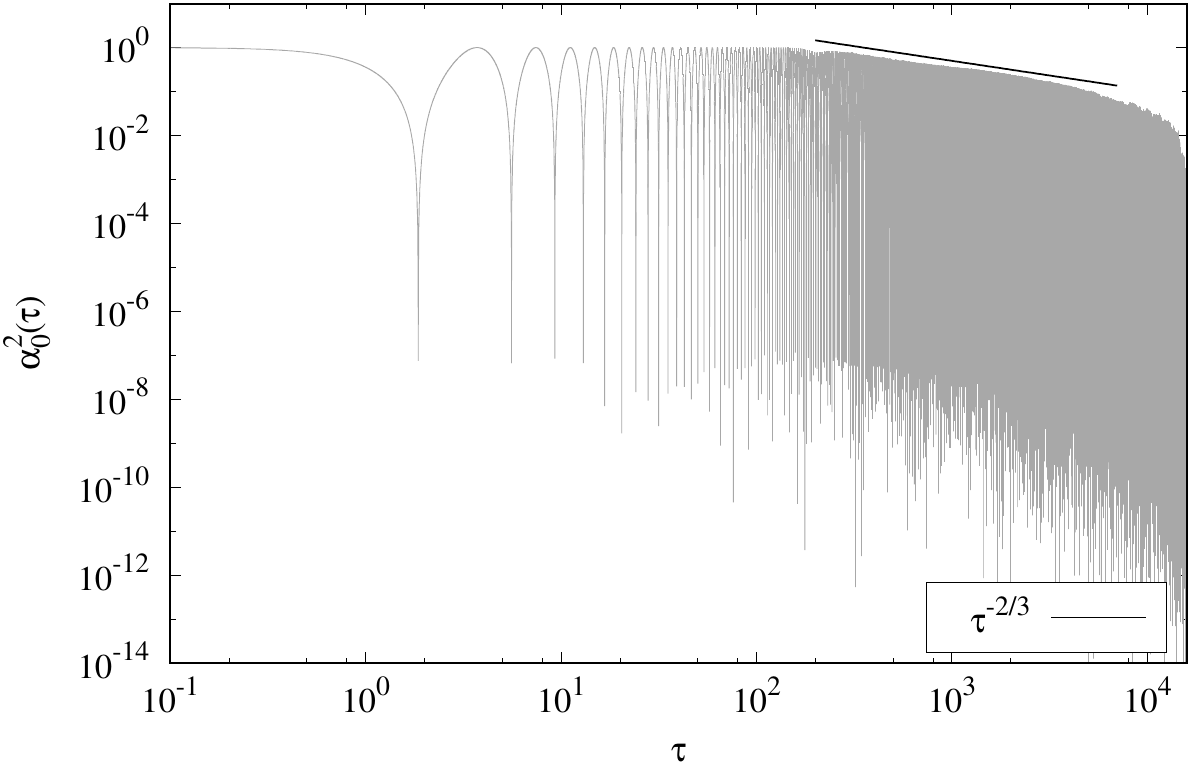}
\caption{\label{hom_1}Time evolution of the homogeneous mode, $\xi = -1$.}
\end{figure}
We notice that for the selected values of $\xi$, the time evolution of $\alpha_\mathbf{0}(\tau)$ and  $\sigma^2_\phi(\tau)$ have a similar structure: the homogeneous mode oscillates with a constant amplitude until $\tau \approx 100$, and the variance (\ref{var_phi}) experiences exponential growth with a resonant peak approximately at the same instant. These aspects characterize the linear stage of the system when some of the inhomogeneous modes growth by the parametric resonance mechanism. From this instant, the nonlinearities of the system play a nonnegligible role leading the system to a stage of energy transfer through different scales called ``driven turbulence'', where $\alpha_\mathbf{0}$ plays the ``energy source'' role. Besides, the amplitude of $\alpha_\mathbf{0}^2$ decays approximately with a power law $\tau^{-2/3}$ and the variance grows with power-law $\tau^{1/5}$, as shown in \cite{micha_03,micha_04}. However, we did not observe the decay of the variance with power-law $\tau^{-2/5}$ as found in \cite{micha_04}, but instead, it seems that the variance reaches an approximate plateau. 
\begin{figure}[h!]
\includegraphics[scale=.6]{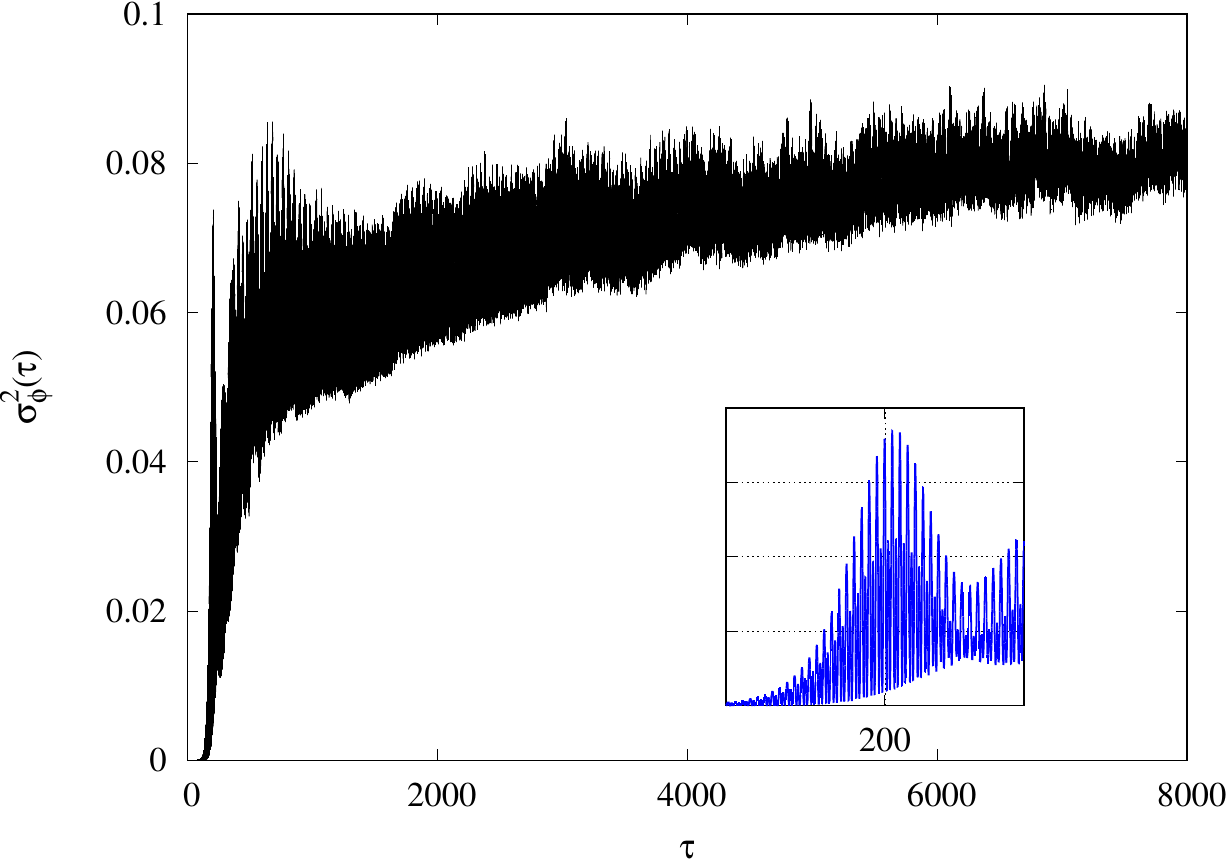}
\includegraphics[scale=.67]{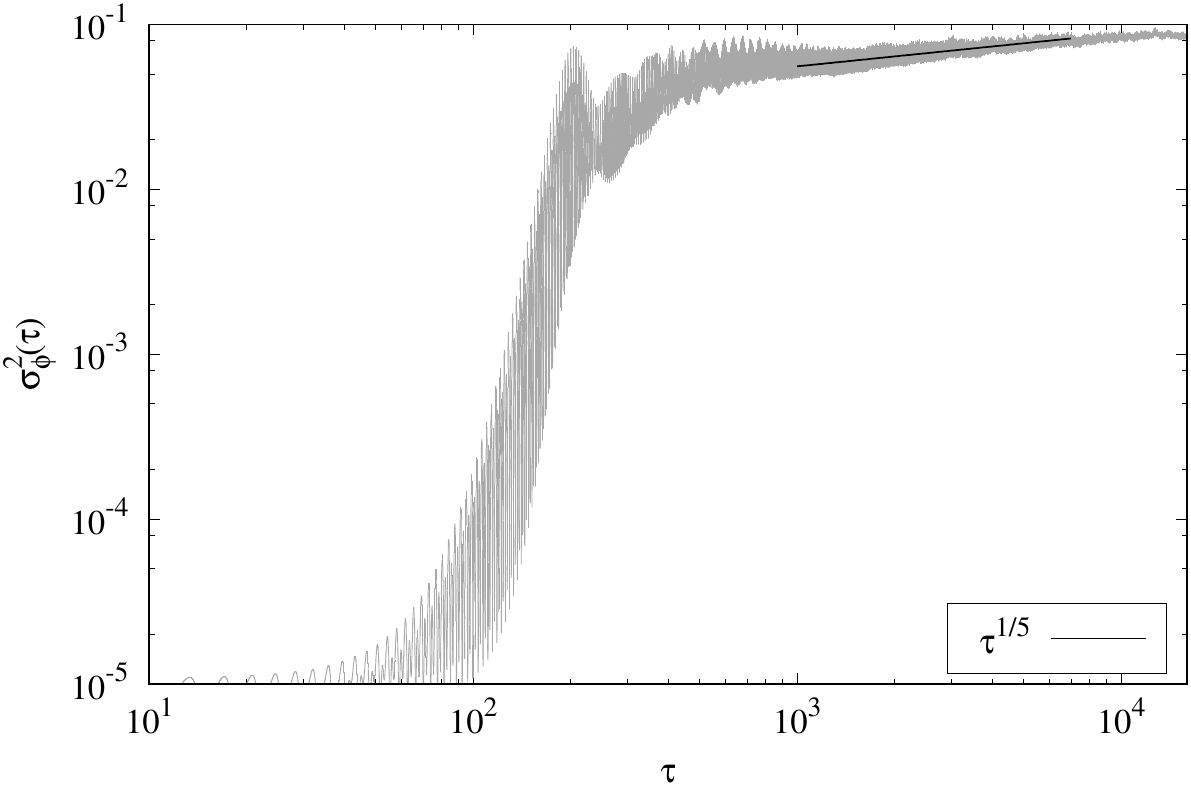}
\caption{\label{var_0} Time evolution of the inflaton variance, $\xi = 0$.}
\end{figure}
\begin{figure}[h!]
\includegraphics[scale=.6]{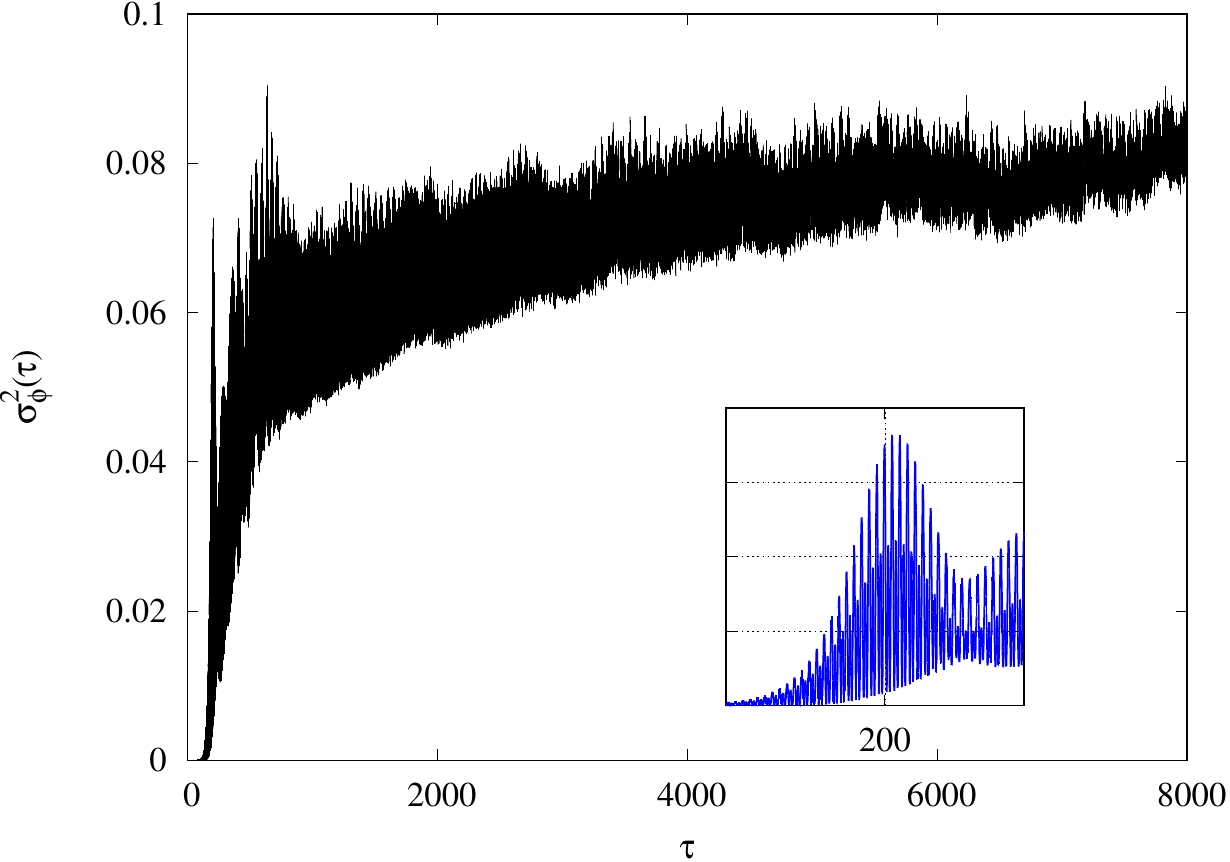}
\includegraphics[scale=.67]{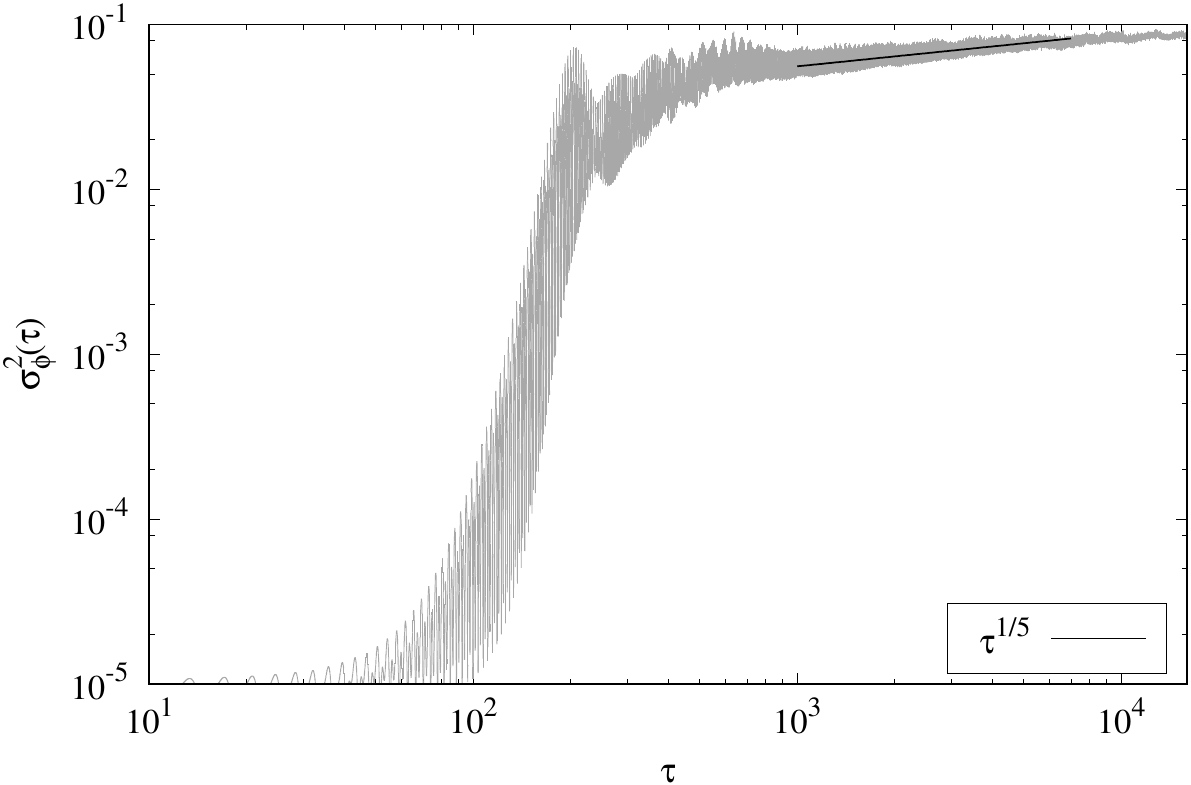}
\caption{\label{var_01} Time evolution of the inflaton variance, $\xi = -0.1$.}
\end{figure}
\begin{figure}[h!]
\includegraphics[scale=.6]{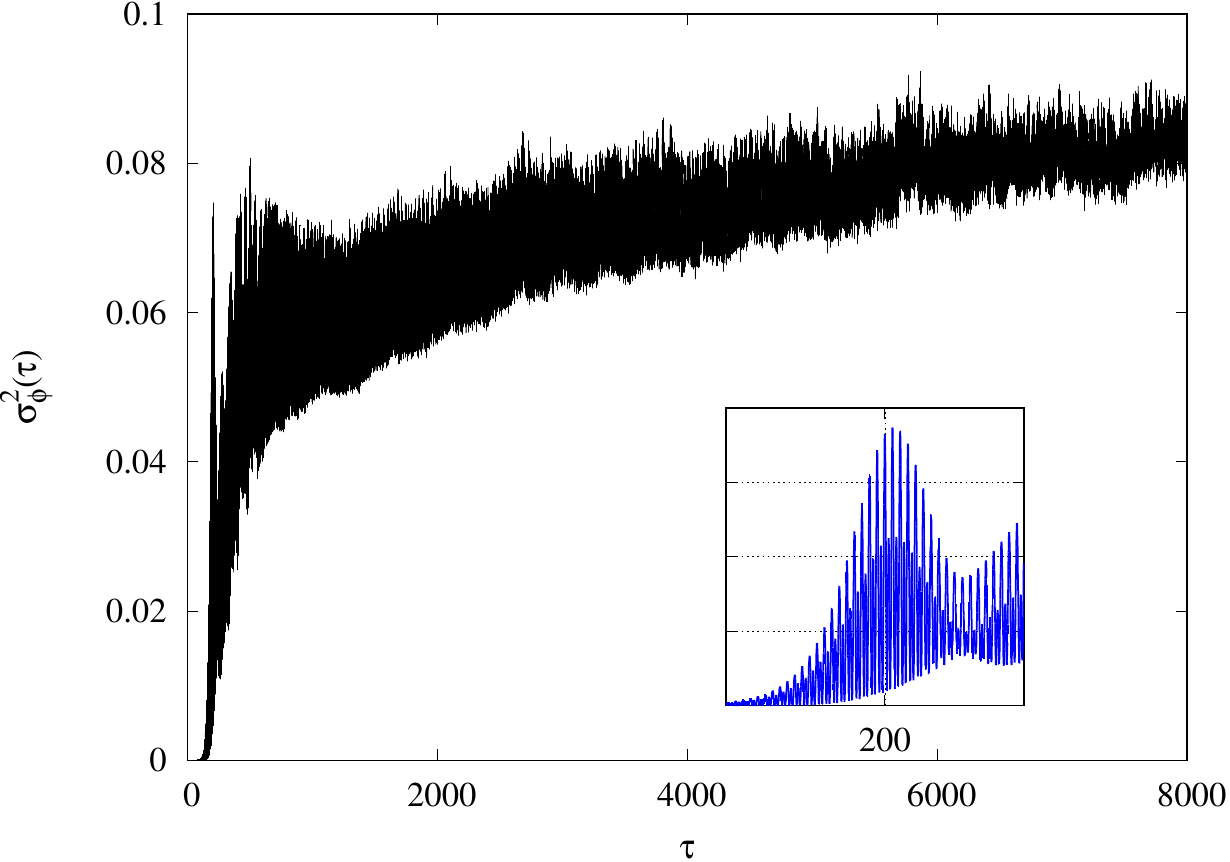}
\includegraphics[scale=.67]{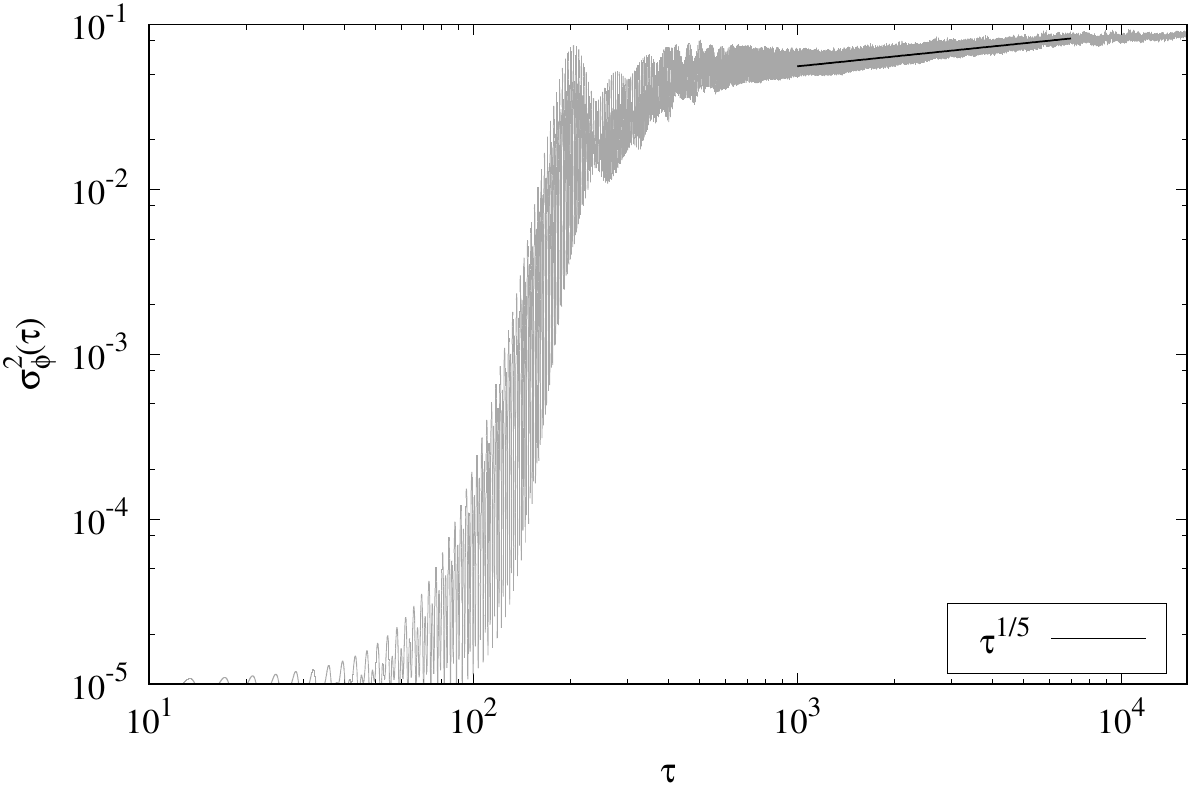}
\caption{\label{var_1} Time evolution of the inflaton variance, $\xi = -1$.}
\end{figure}

\begin{figure}[h!]
\begin{center}
\includegraphics[scale=.6]{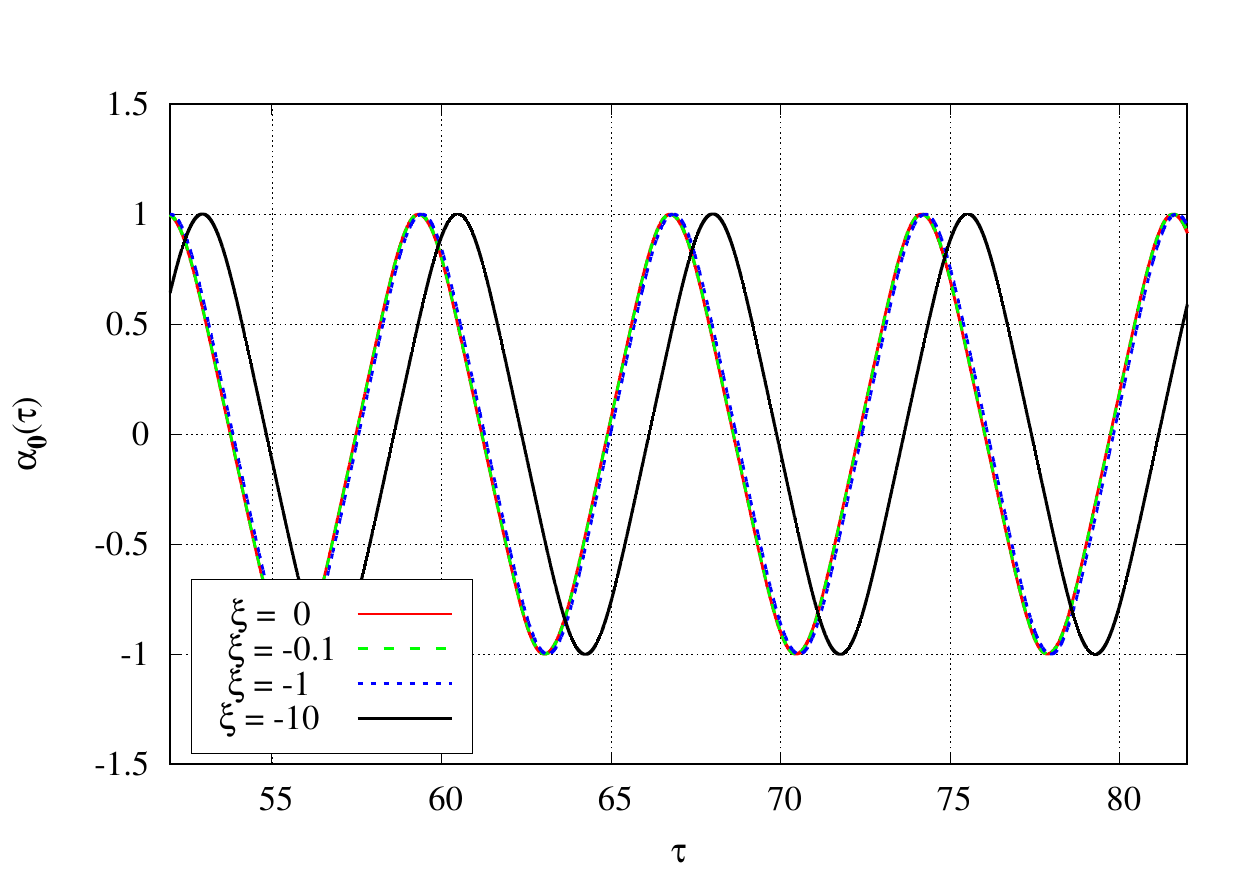}
\includegraphics[scale=.6]{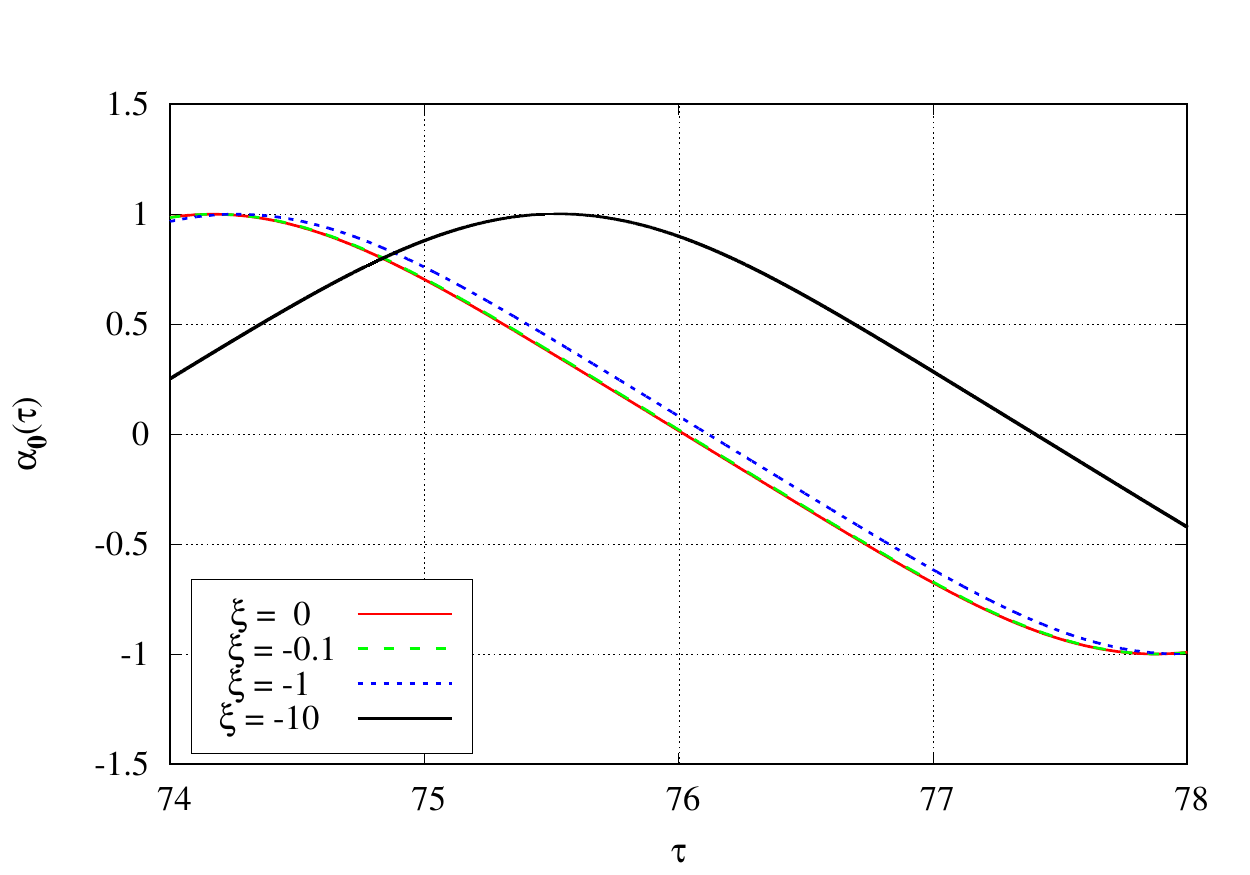}
\end{center}
\caption{\label{hom_comp}Influence of the value of $\xi$ on the natural frequency of $\alpha_{\mathbf{0}}\left(\tau\right)$.}
\end{figure}

According to Figs. \ref{hom_0} - \ref{var_1}, the dynamical behavior of these models is qualitatively similar. We noticed that the influence of the nonminimal coupling in the linear stage is to decrease the natural frequency of the homogeneous model slightly as illustrated in Fig. \ref{hom_comp}. The periods for $\xi=0, -0.1, -1.0$  and $-10.0$ are approximately $\Pi_\phi \approx$ 7.19, 7.20, 7.36 and 7.597, respectively. Therefore, for a small value of $|\xi|$ the behavior of the variance is not altered substantially, contrary to the results presented by Tsujikawa et al. \cite{tsujikawa}.

\subsection{Back reaction effects}

We proceed with the influence of the nonminimal coupling parameter $\xi$ on the backreaction of the created particles in the effective equation of state and the expansion of the Universe described by the scale factor $a(\tau)$. The effective equation of state is denoted by the parameter $w$ given by

\begin{equation}
w\left(\tau\right) = \dfrac{\left\langle p \right\rangle}{\left\langle \rho \right\rangle}, \label{eos}
\end{equation}
\noindent where $\left< p \right>$ and $\left< \rho \right>$ are the effective  pressure and energy density, respectively with the corresponding expressions derived from the field equations (\ref{einstein_eq}):

\begin{eqnarray}
\rho & = & \dfrac{\lambda\phi_0^4}{a^4\left(1-\xi\phi_0^2 m_{pl}^{-2}a^{-2}\phi_p^2\right)}\left\lbrace \dfrac{1}{2}\left(\phi'_p - H_p \phi_p\right)^2 + \left(\dfrac{1}{2} - 2\xi\right)\left(\boldsymbol{\nabla}_p\phi_p\right)^2 + \dfrac{1}{4}\phi_p^4 + \right. \nonumber\\
      &    &\  \left. + 2\xi\phi_p\left[3H_p\left(\phi'_p - H_p\phi_p\right) - \nabla_p^2\phi_p\right]\right\rbrace, \\
     \nonumber \\
p    & = &  \dfrac{\lambda\phi_0^4}{a^4\left(1-\xi\phi_0^2 m_{pl}^{-2}a^{-2}\phi_p^2\right)}\left\lbrace \left(\dfrac{1}{2}-2\xi\right)\left(\phi'_p - H_p \phi_p\right)^2 - \left(\dfrac{1-8\xi}{6}\right)\left(\boldsymbol{\nabla}_p\phi_p\right)^2 - \right. \nonumber\\
     &   & \left. - \dfrac{1}{4}\phi_p^4 + 2\xi\phi_p\left[H_p\phi'_p - \left(H_p^2-6\xi\dfrac{a''}{a}\right) \phi_p -\dfrac{1}{3} \nabla_p^2\phi_p +\dfrac{1}{4}\phi_p^4\right]\right\rbrace,
\end{eqnarray}

\begin{figure}[h!]
\includegraphics[scale=.63]{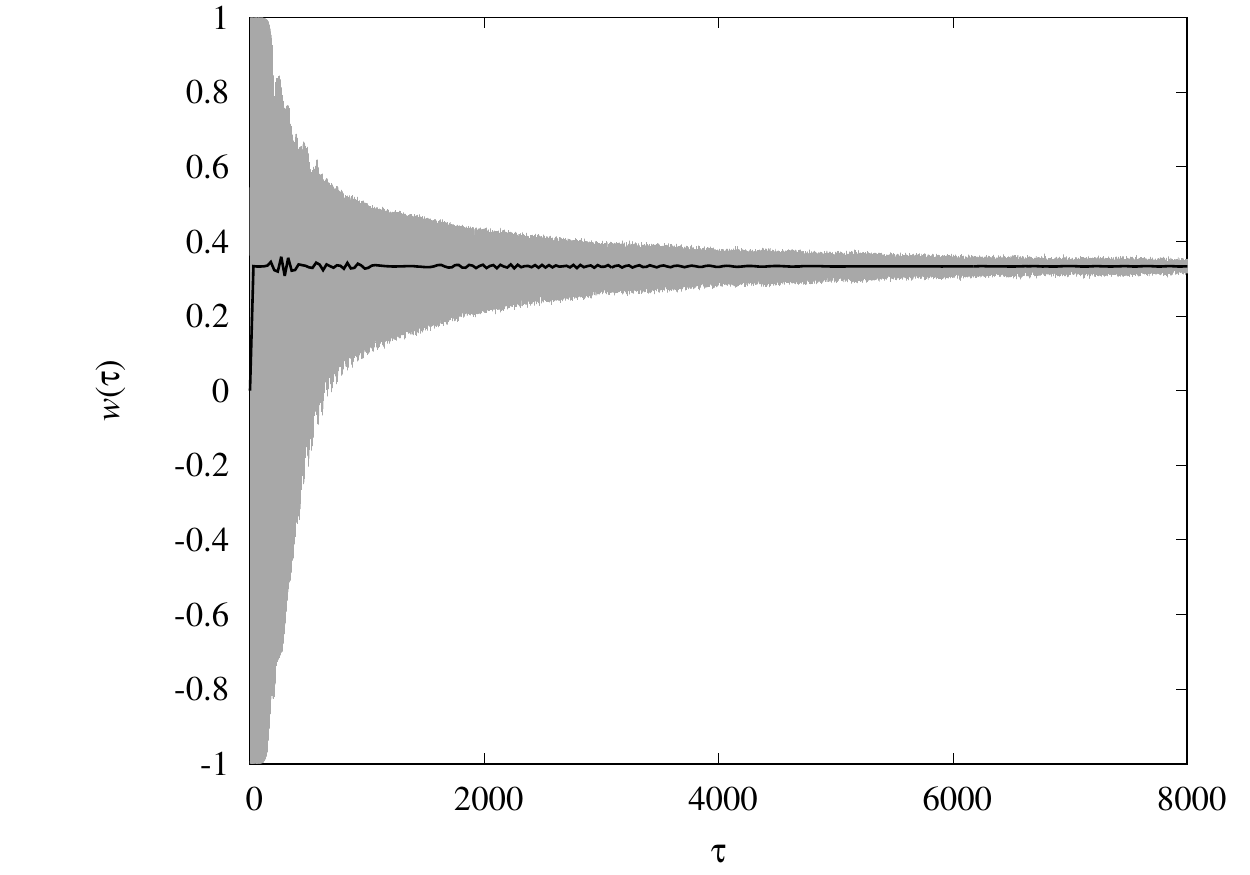}
\includegraphics[scale=.63]{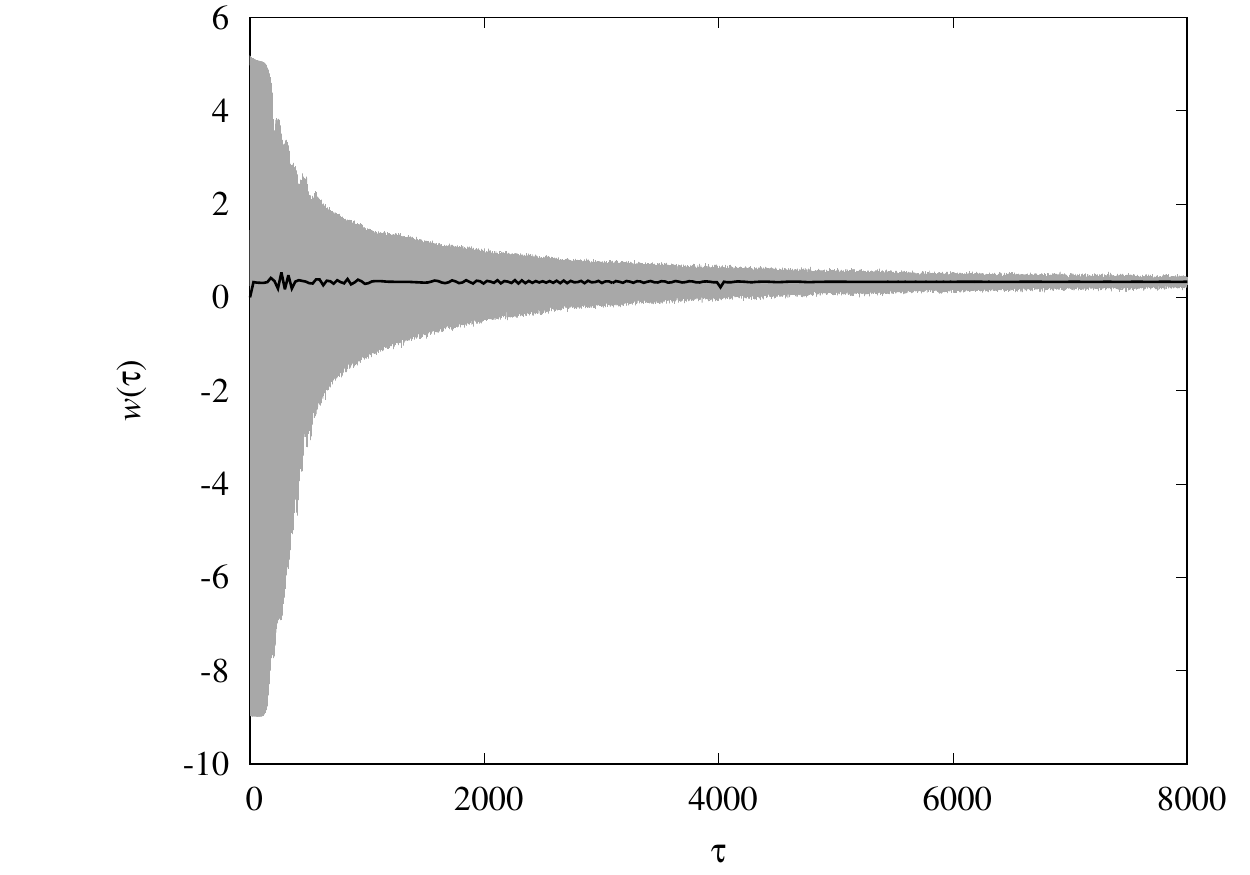}
\caption{\label{br_0}$w$ time evoltution for $\xi = 0$ and $\xi = -1$.}
\end{figure}

\begin{figure}[h]
\begin{center}
\includegraphics[scale=.61]{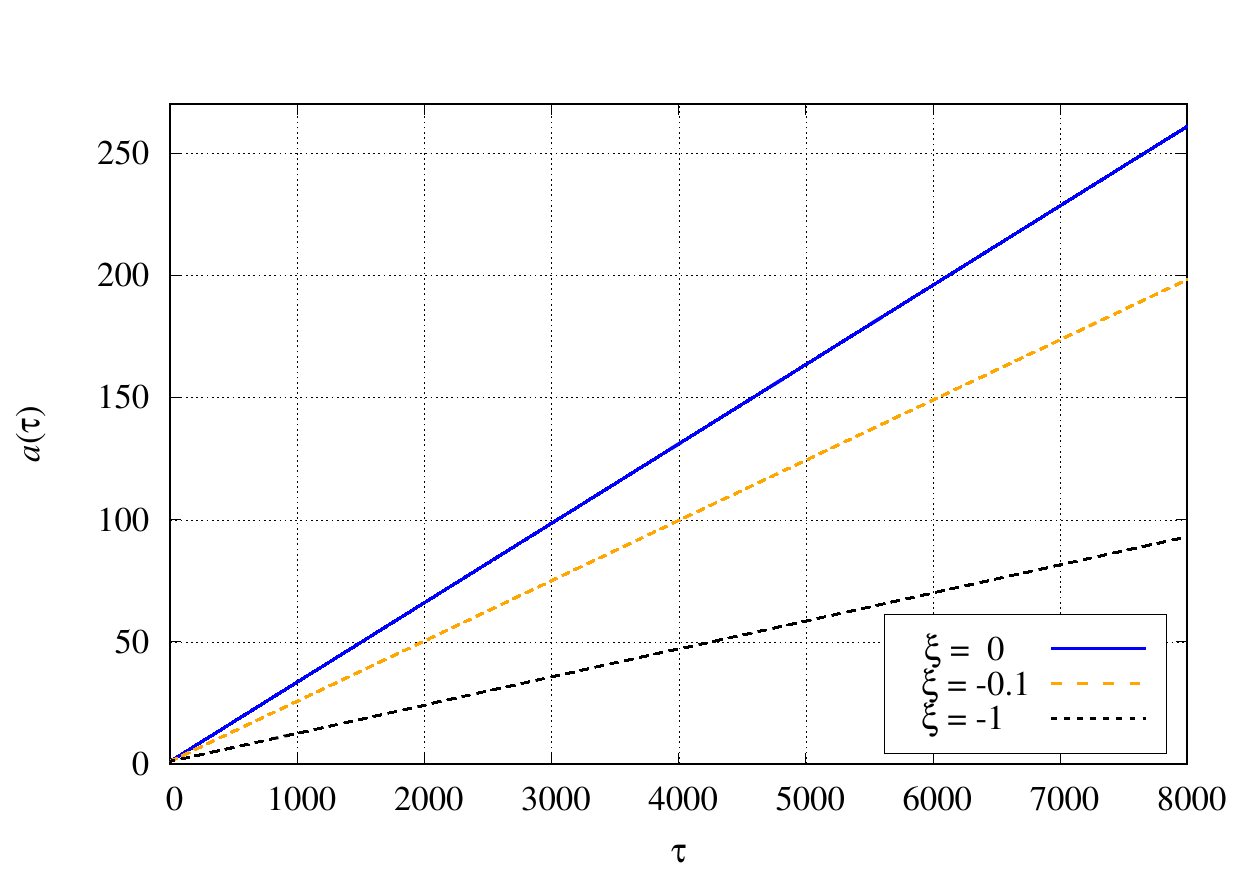}
\includegraphics[scale=.61]{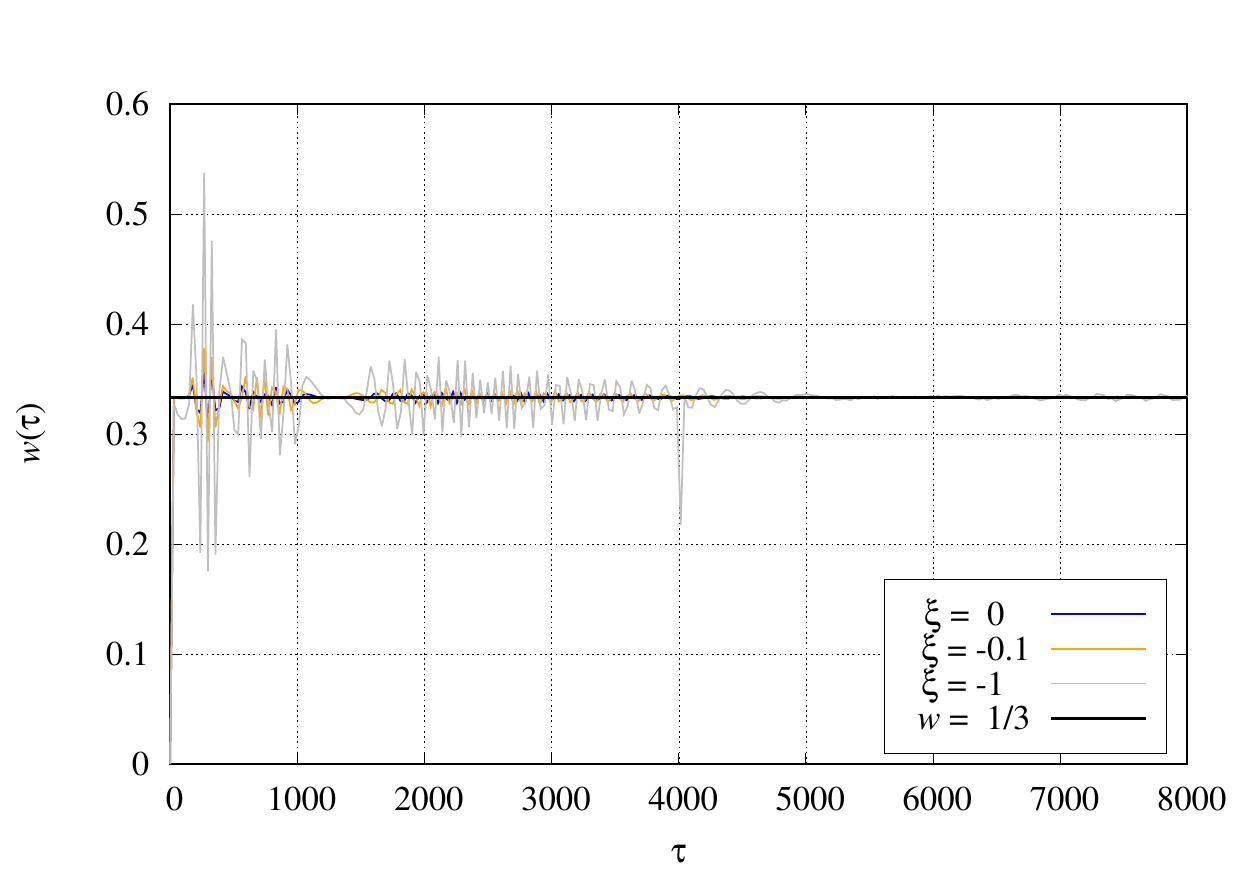}
\end{center}
\caption{\label{w_comp} Mean behavior of $a\left(\tau\right)$ and $w\left(\tau\right)$ for different values of $\xi$.}
\end{figure}

We show in Fig. \ref{br_0} the time evolutions of the effective equation of state $w$ for $\xi=0$ (left) and $\xi=-1$ (right). The mean behavior of $w$ tends to  $w\approx 1/3$ (black line in the right side of Fig. \ref{w_comp}) for all values $\xi$ that we have considered. This value would be expected for the radiation era. The noticeable effect of the parameter $\xi$ is to change the speed of growth of the scale factor $a(\tau)$ (left side of Fig. \ref{w_comp}) but since the oscillations of the homogeneous component is not drastically affected, the energy transfer mechanism seems is almost the same for small values of $\xi$. The same occurs with the effective equation of state whose describing a radiation fluid in average.

\subsection{Power spectra in spatial domain}

Power spectra in the spatial domain (wave vectors) provide essential aspects about the energy transfer through different scales of the system in the turbulent phase with the homogeneous mode $(k \equiv \left\vert \mathbf{k} \right\vert = 0$) acting as the energy source. The energy flows from larger scales, represented by the modes with small $k$, to smaller scales represented by modes with greater $k$, through constant fluxes named energy cascades. The presence of these cascades is signalized by scale laws on the power spectra in the spatial domain. In particular, for the hydrodynamic turbulence, some intervals of these power spectra should be described by power laws like 
\begin{equation}
P\left(k\right) \propto k^{-\gamma}. \label{kolmogorov_law}
\end{equation}

To analyze the features of the wave turbulence, we have constructed the power spectra of two quantities related to the system. The first one is the function $\textrm{var}\left(\phi_p\right)$ given by
\begin{equation}
\textrm{var}\left(\phi_p\right) = \left(\phi_p - \left\langle\phi_p\right\rangle\right)^2. \label{var_fc}
\end{equation}

\noindent And the second  function is the rescaled energy density $\bar{\rho}$
\begin{equation}
\bar{\rho} = \dfrac{a^4\rho}{\lambda\phi_0^4}.
\end{equation}

\noindent In both cases we evaluated the expansion of these quantities in terms of Fourier basis: 
\begin{eqnarray}
\textrm{var}\left(\phi_p\right) & = & \sum\limits_{k_x,k_y,k_z = -N_e/2}^{N_e/2-1} \widehat{v}_\mathbf{k}\expt^{2\pi i \mathbf{k}\cdot\mathbf{x}_p/L_p}, \\
\bar{\rho} & = & \sum\limits_{k_x,k_y,k_z = -N_e/2}^{N_e/2-1} \widehat{e}_\mathbf{k}\expt^{2\pi i \mathbf{k}\cdot\mathbf{x}_p/L_p},
\end{eqnarray}
\noindent where we set $N_e = 8N$ and evaluated these expansions with the same FFT algorithm employed for the spectral expansion of $\phi_p$. The power spectra $P_\sigma\left(k\right)$ and $P_\rho\left(k\right)$ were obtained calculating the the root mean squares of the modes  $\left\vert\widehat{v}_\mathbf{k}\right\vert$ and $\left\vert\widehat{e}_\mathbf{k}\right\vert$, respectively, having the same $k \equiv \left\vert \mathbf{k} \right\vert$. In particular, $\widehat{v}_\mathbf{0} = \sigma^2_\phi$ (cf. Eq. (\ref{var_phi})).

\begin{figure}[h!]
\includegraphics[scale=0.61]{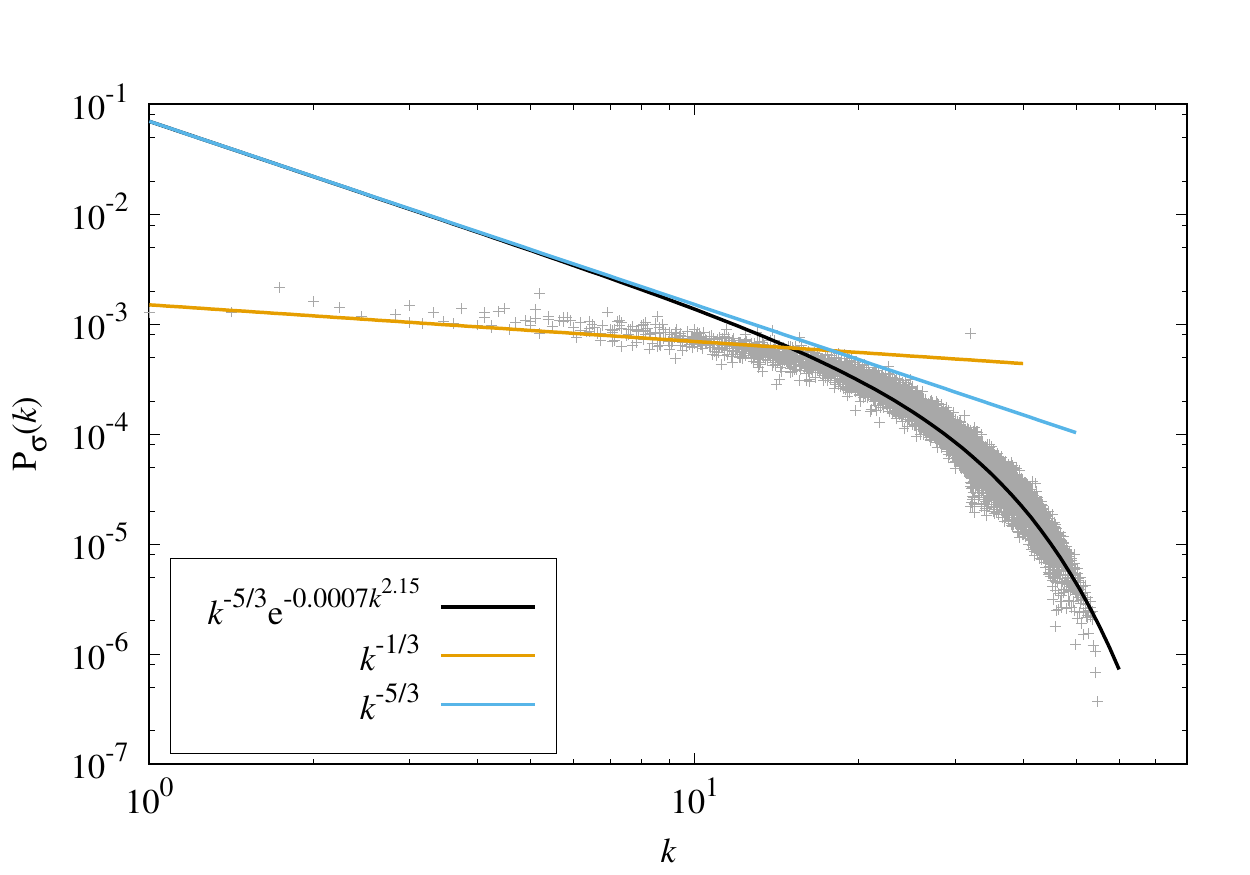}
\includegraphics[scale=0.61]{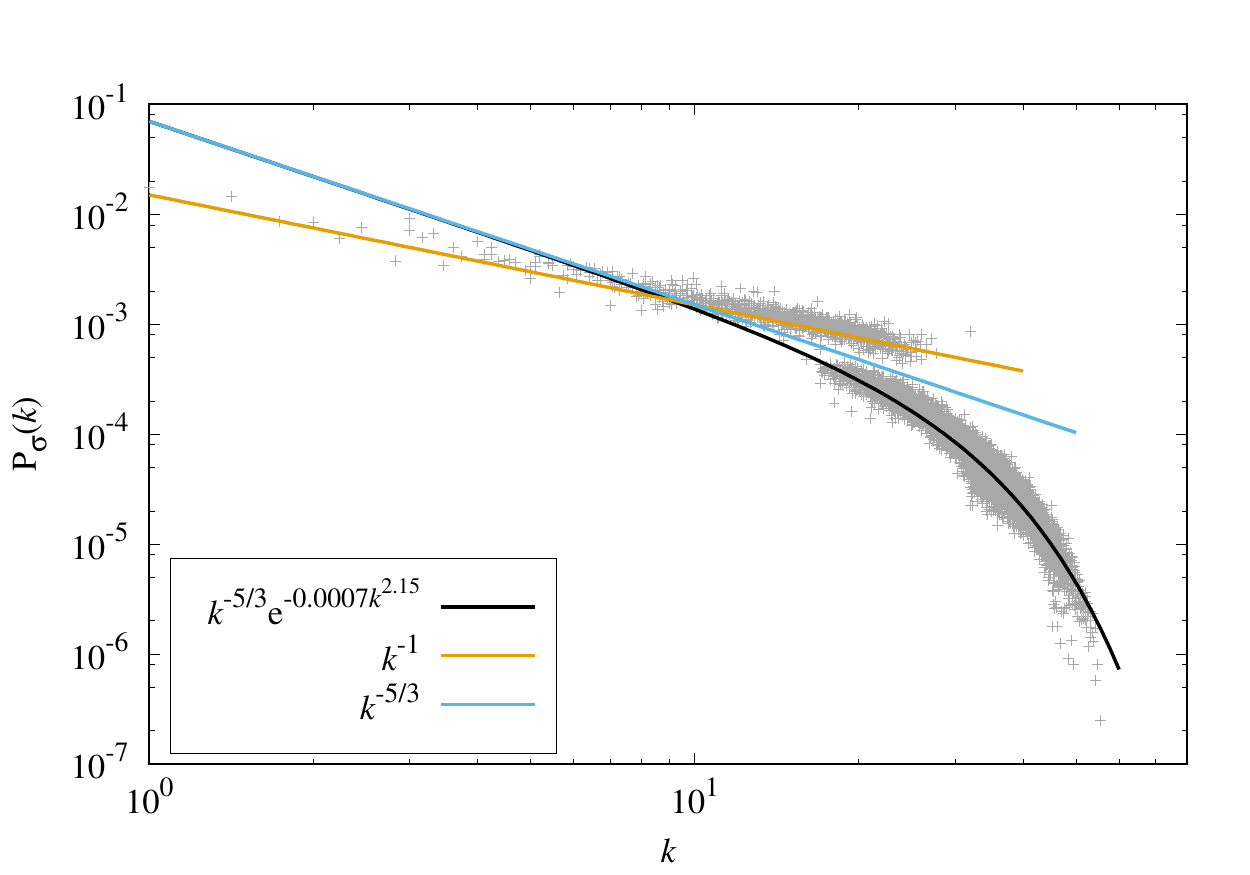}
\caption{\label{ps_0} Variance power spectra in spatial domain for $\xi = 0$ evaluated at $\tau \approx 8380$ and $\tau \approx 16770$.}
\end{figure}

\begin{figure}[h!]
\includegraphics[scale=0.61]{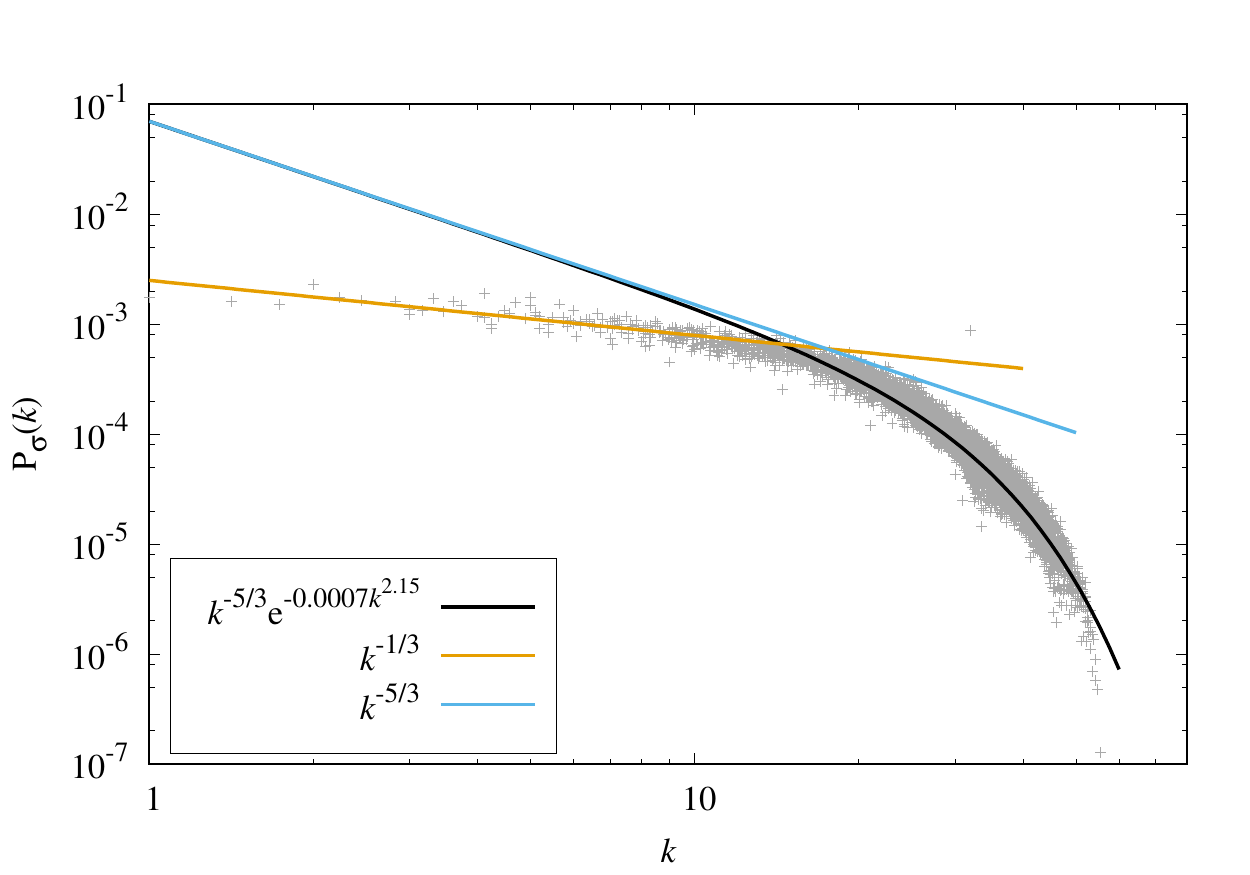}
\includegraphics[scale=0.61]{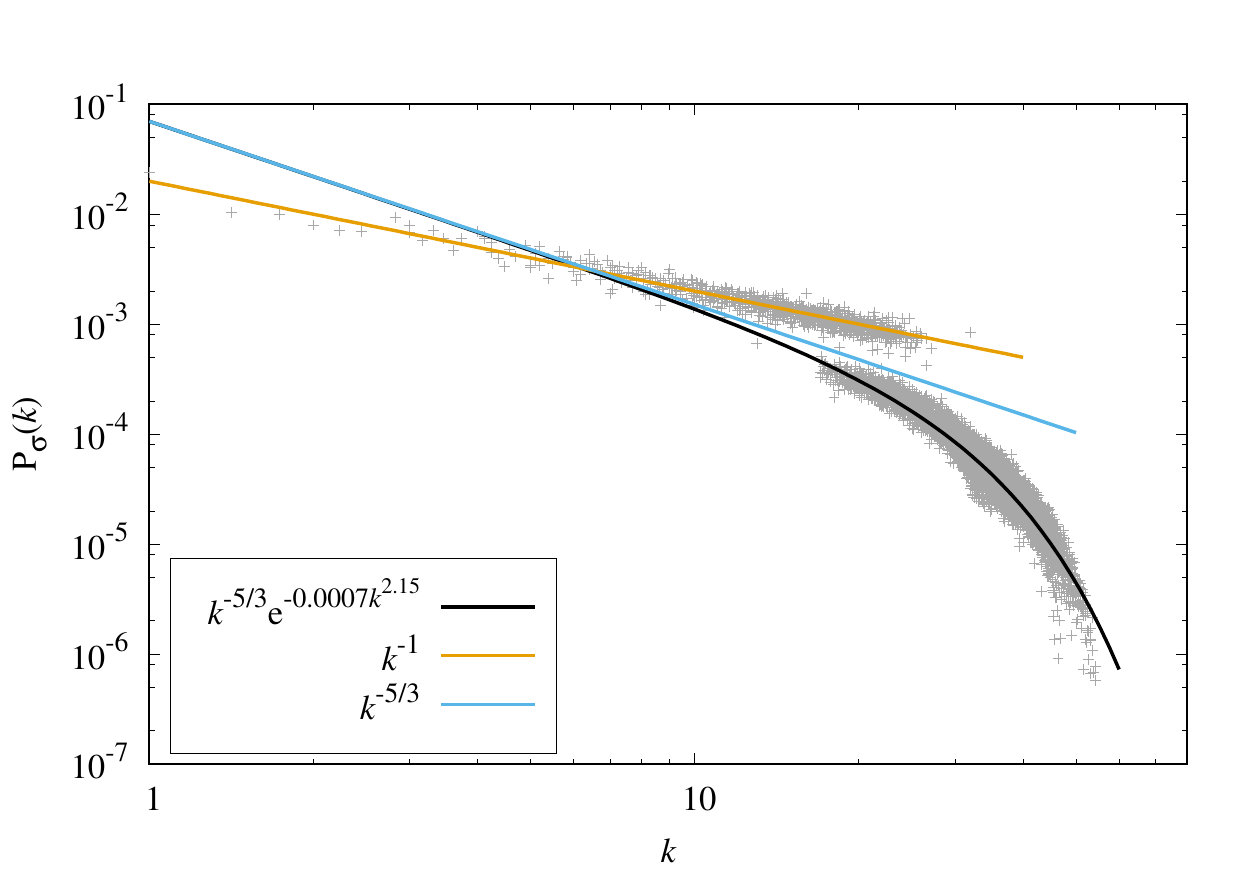}
\caption{\label{ps_01} Variance power spectra in spatial domain for $\xi = -0.1$ evaluated at $\tau \approx 8380$ and $\tau \approx 16770$.}
\end{figure}

\begin{figure}[h!]
\includegraphics[scale=0.61]{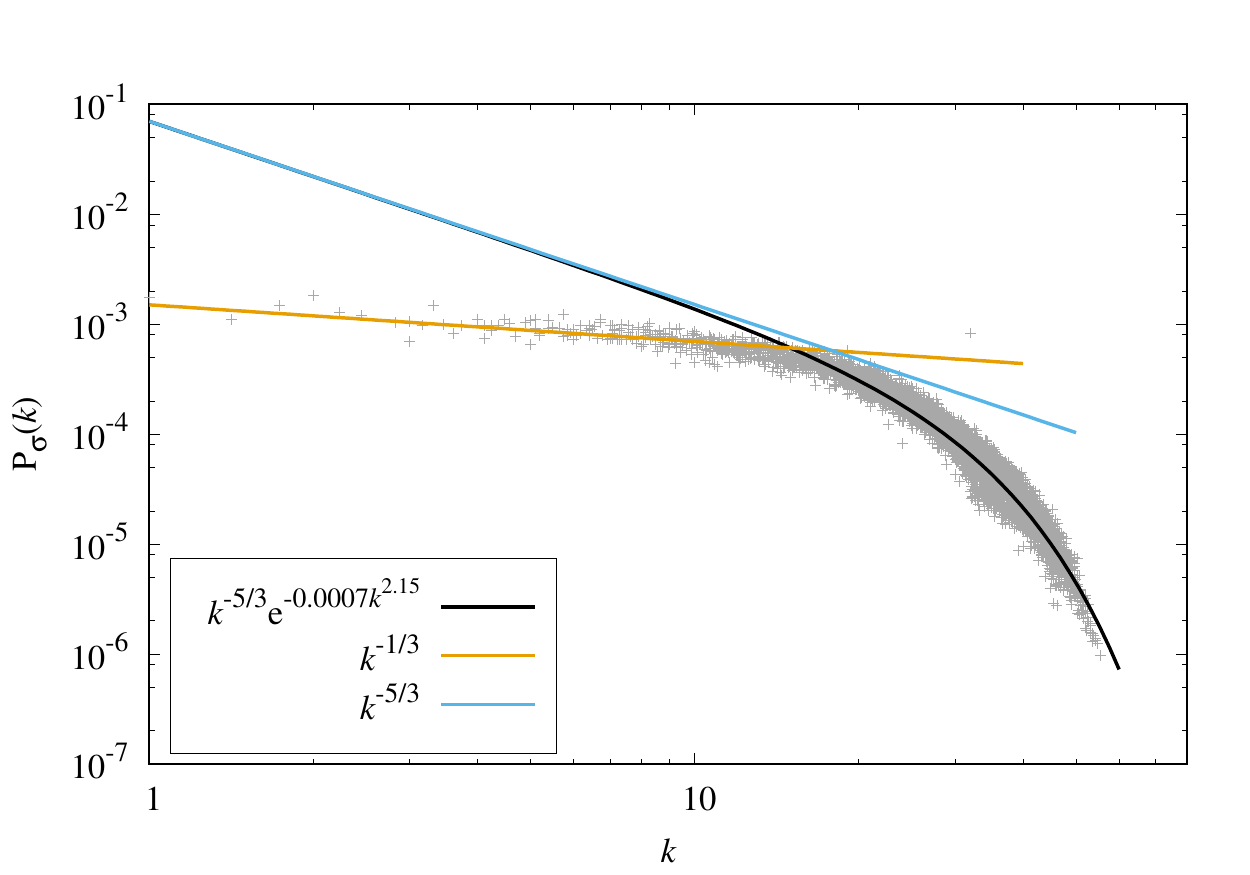}
\includegraphics[scale=0.61]{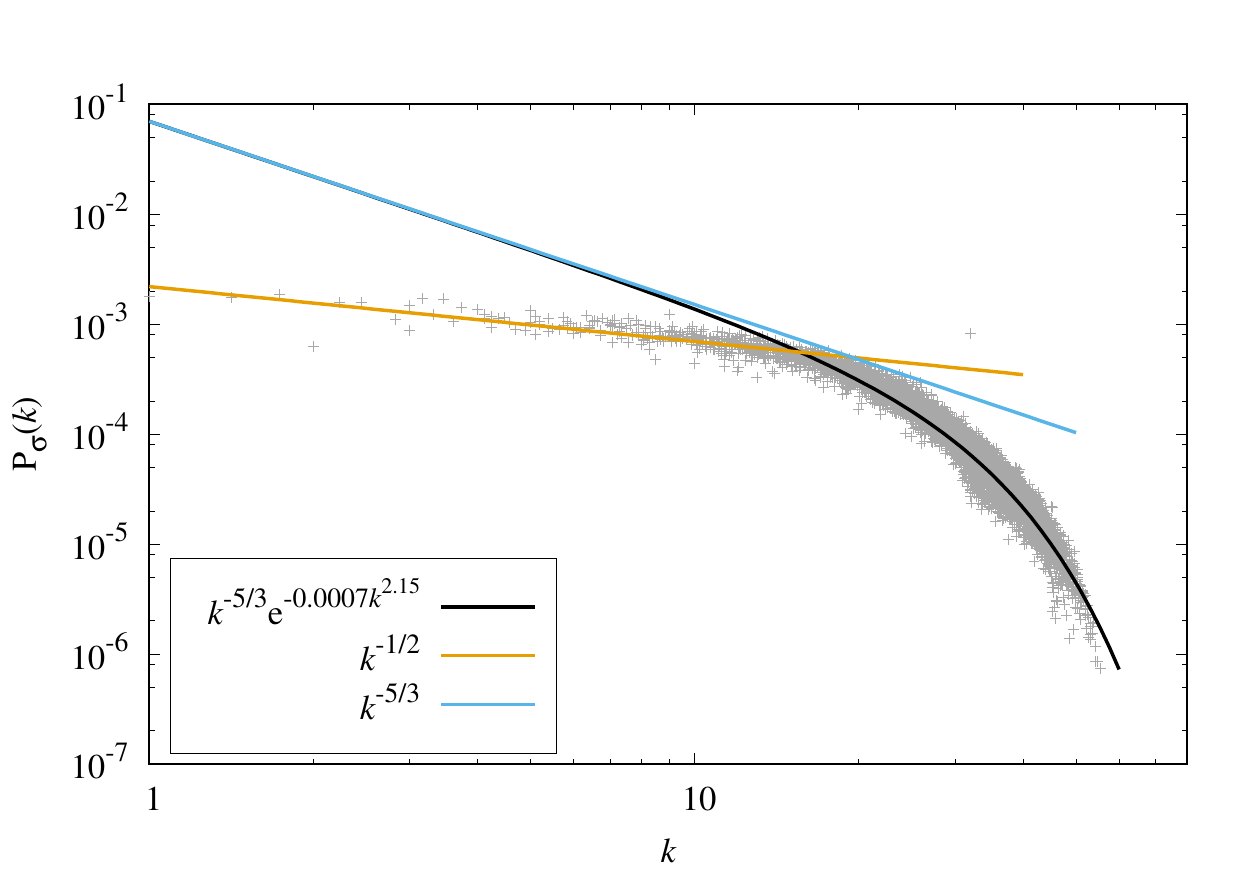}
\caption{\label{ps_1} Variance power spectra in spatial domain for $\xi = -1$ evaluated at $\tau \approx 8380$ and $\tau \approx 16770$.}
\end{figure}

We show in Figs. \ref{ps_0} - \ref{ps_1} the power spectra of $\textrm{var}\left(\phi_p\right)$ and $\bar{\rho}$ evaluated at $\tau \approx 8380$ (panels on the left) and at $\tau \approx 16770$ (panels on the right) for $\xi=0,\xi=-0.1$ and $-1$, respectively. We noticed that depending on the range of $k$ they share the same scale law. In the wavenumber interval $0 \leqslant k \apprle 29$ we found $P_\sigma \simeq k^{-\gamma}$ with $\gamma \approx 1/3$ at $\tau = 8380$ for all values of the parameter $\xi$; at $\tau = 16770$, the exponent $\gamma$ changes according to $\xi$ as: $\gamma \approx 1$ for $\xi= 0$ and $-0.1$, $\gamma \approx 1/2$ for $\xi=-1$ and $\gamma \approx 3/4$ for $\xi=-10$. More interestingly, in the interval $29 \lesssim k \leqslant k_{\mathrm{max}}=80$, the scale law 
\begin{equation}
P_\sigma\left(k\right) \propto k^{-5/3}\exp^{-7.0 \times 10^{-4} k^{2.15}},
\end{equation}

\noindent is valid for all values of the parameter $\xi$ and in both instants under consideration. In particular, this type of spectrum decay in wave numbers occurs in magnetohydrodynamic turbulence \cite{terry_09} and reflects a Kolmogorov-like law due to the turbulent cascade. 

The power spectrum of the energy exhibits a more complex structure than the corresponding to the variance. For $k \lesssim 10$, all the spectra are described by the power-law  (\ref{kolmogorov_law}) with $\gamma \simeq 1/7$. For $k > 10$ we were able to identify two distinct components described by the following one-parameter function
\begin{equation}
P_\rho\left(k\right) \propto \dfrac{k^3}{\expt^{bk}-1},\label{ps_rho}
\end{equation}

\noindent where $b \approx 0.18$ is independent of $\xi$ as well the instants of time under consideration. We identify this scaling law as the Planck's law that describes the radiation spectrum of a black body. It might be remarked that the value of $b$ found in the present three-dimensional model is the same we have obtained in a two-dimensional model \cite{crespo_14} of preheating after inflation with a minimally coupled inflaton field.

\begin{figure}[h!]
\includegraphics[scale=0.61]{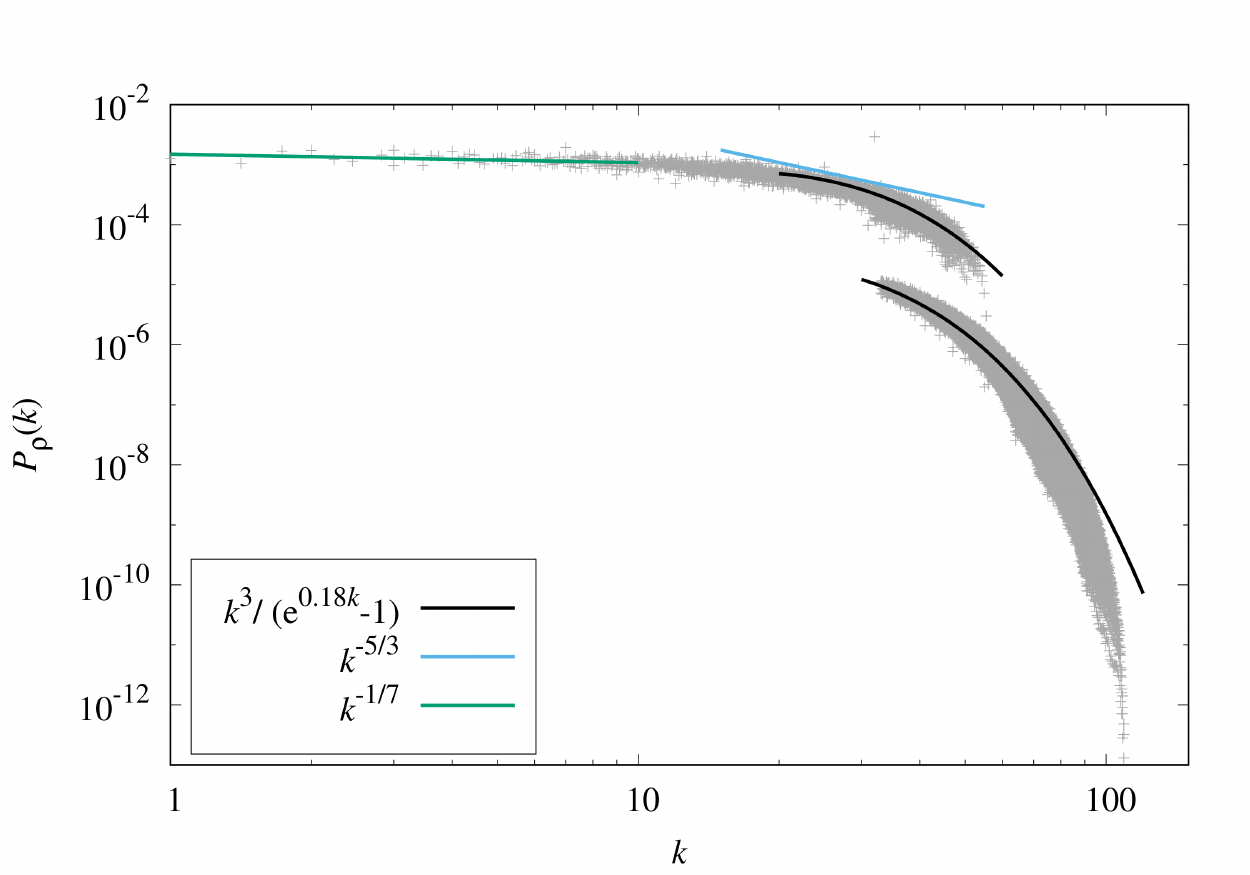}
\includegraphics[scale=0.61]{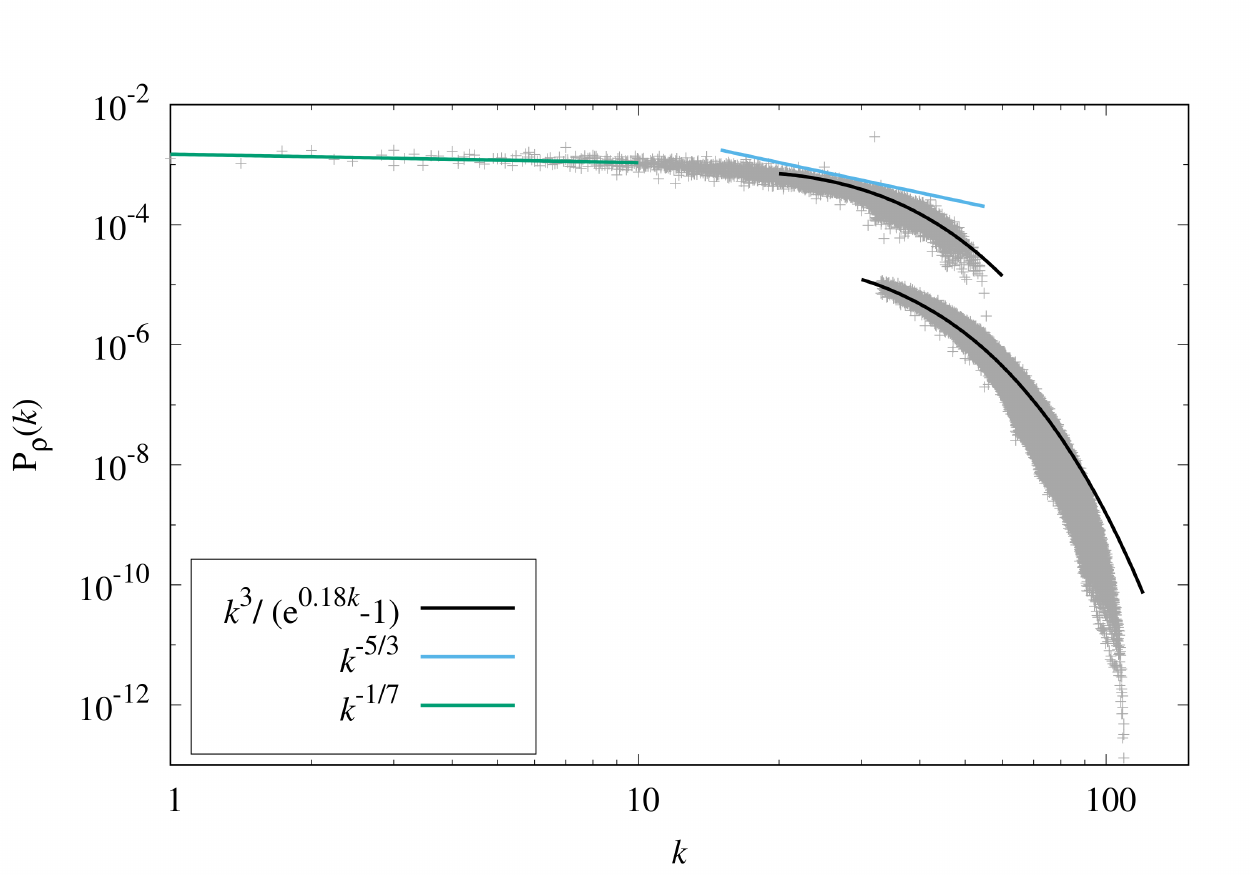}
\caption{\label{ps_0a} Power spectra in spatial domain for $\xi = 0$ evaluated at $\tau \approx 8380$ and $\tau \approx 16770$.}
\end{figure}

\begin{figure}[h!]
\includegraphics[scale=0.61]{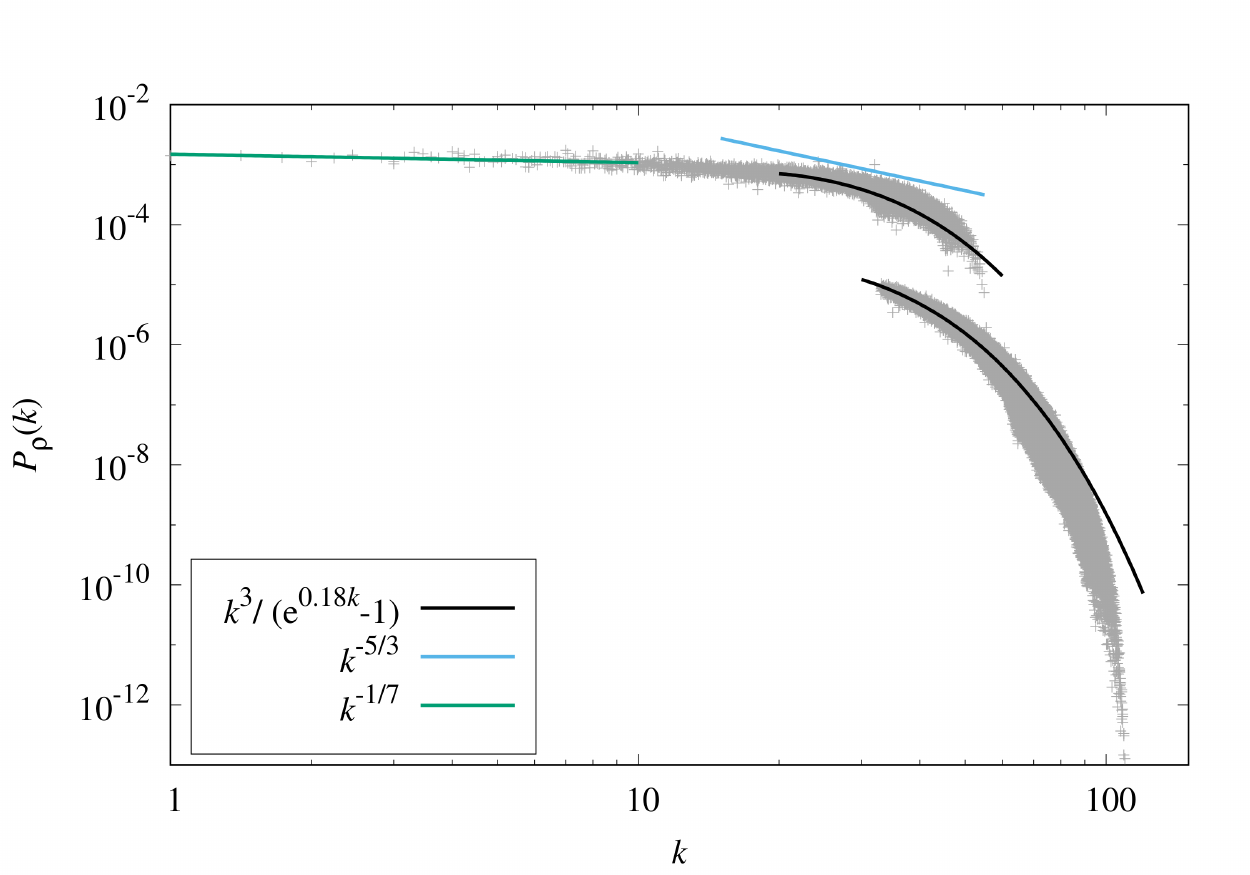}
\includegraphics[scale=0.61]{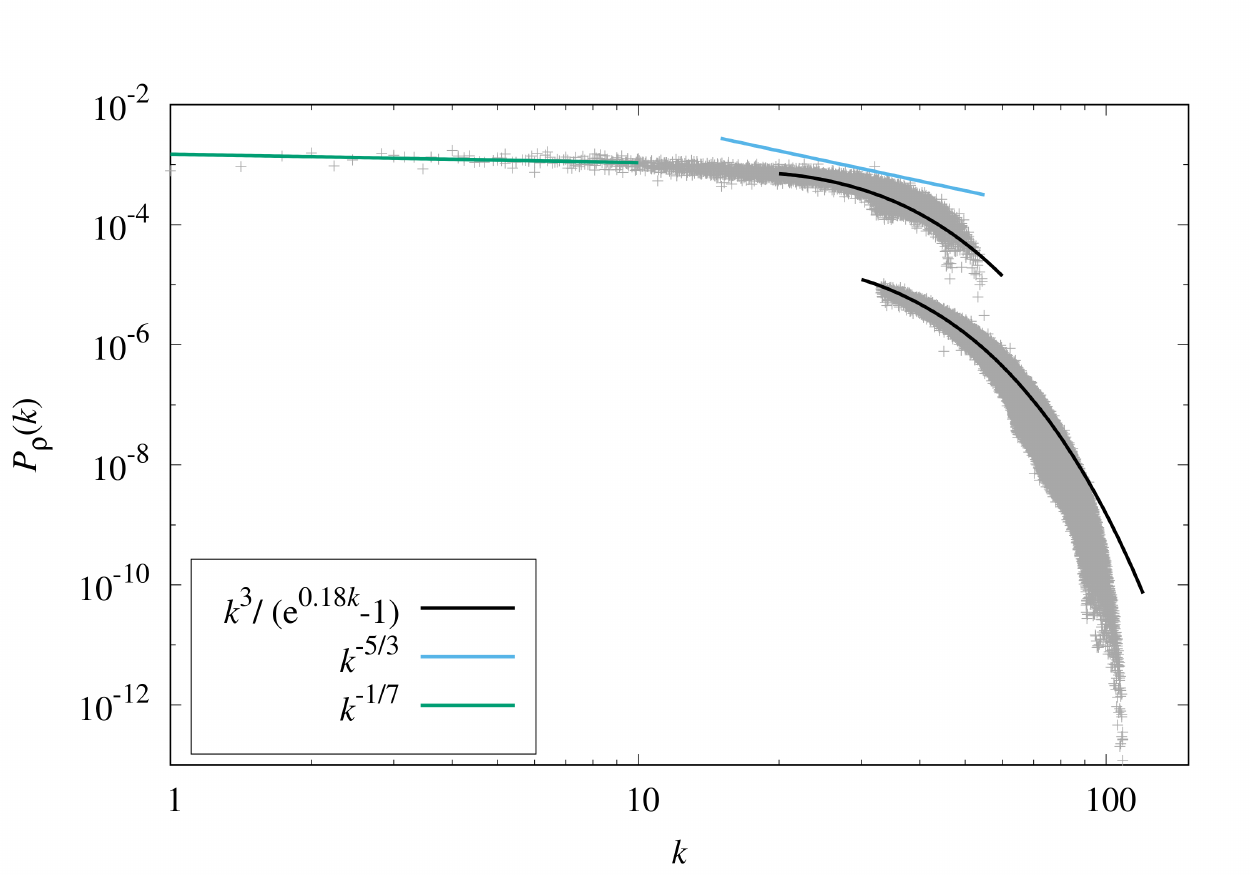}
\caption{\label{ps_01a} Power spectra in spatial domain for $\xi = -0.1$ evaluated at $\tau \approx 8380$ and $\tau \approx 16770$.}
\end{figure}

\begin{figure}[h!]
\includegraphics[scale=0.61]{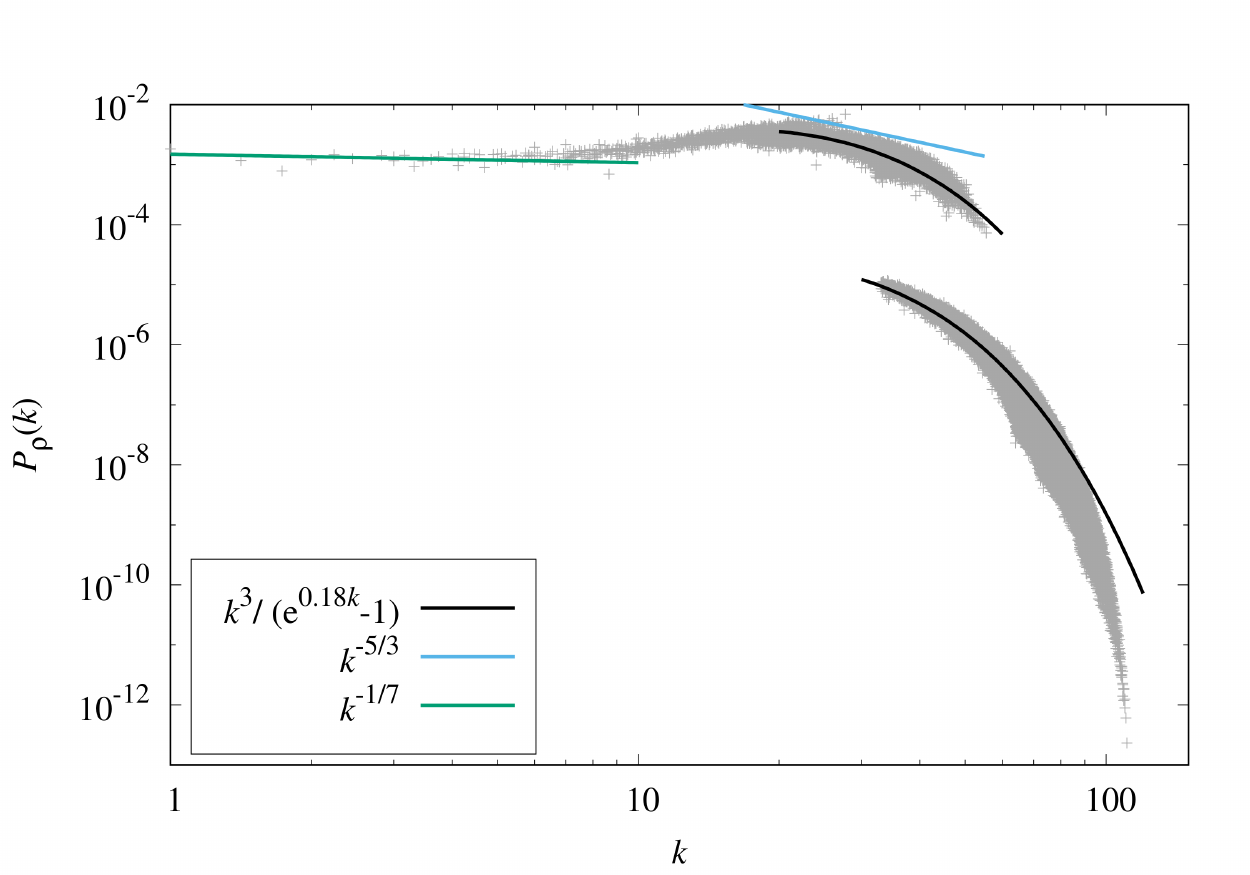}
\includegraphics[scale=0.61]{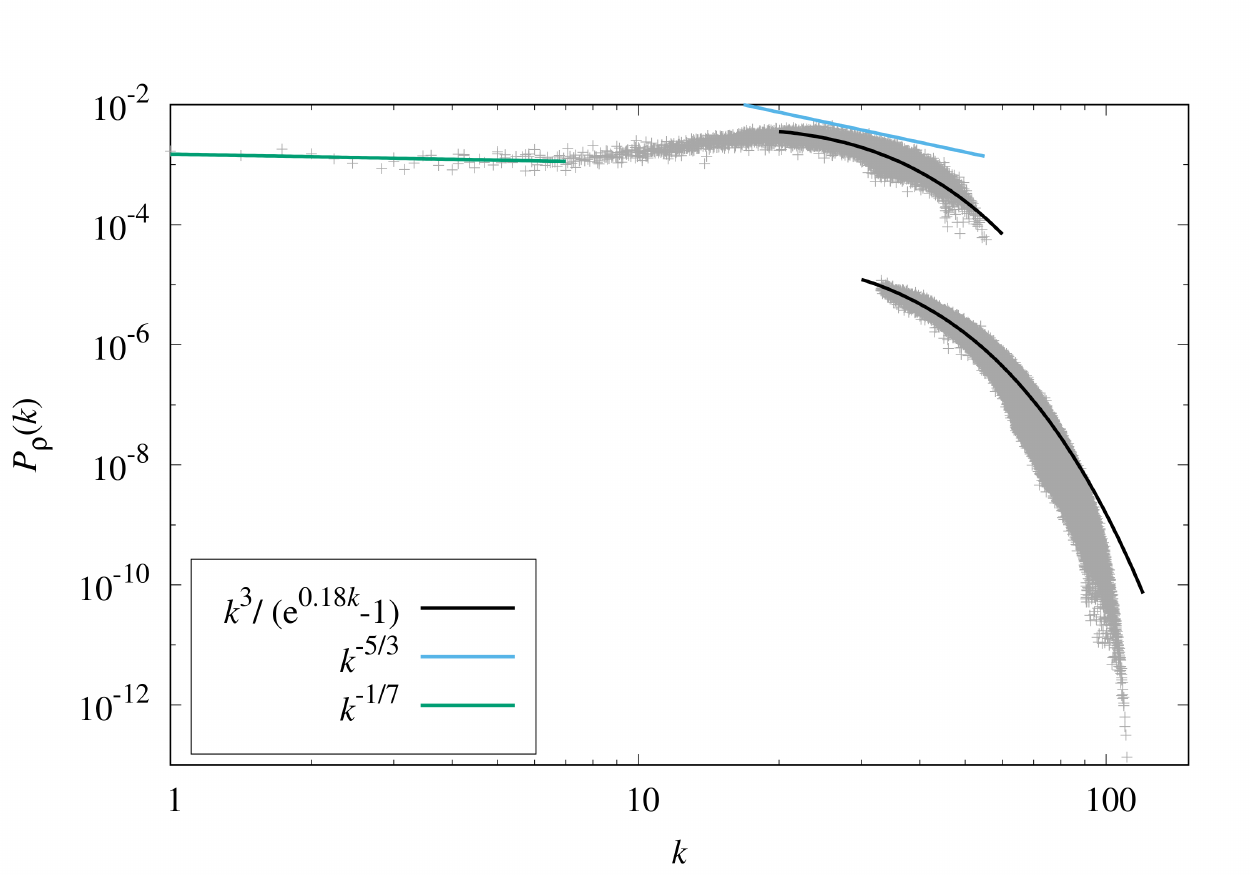}
\caption{\label{ps_1a} Energy power spectra in spatial domain for $\xi = -1$ evaluated at $\tau \approx 8380$ and $\tau \approx 16770$.}
\end{figure}

The scaling law (\ref{ps_rho}) indicates the presence of the energy cascade that drives the system to the typical radiation energy distribution. As a consequence, the wave turbulence is the primary process driving the Universe to the radiation era. This feature is not altered by the values of the parameter $\xi$ despite the details of the power spectrum are slightly altered as it can be verified in Figs. \ref{ps_0a} - \ref{ps_1a}. Another aspect denoting the cascade of energy the presence of an intermediate range that can be described approximately by the power law (cf. eq. \ref{kolmogorov_law}) with $\gamma \approx 5/3$ (blue lines) that reveals the so-called inertial range. In the present simulations, we found the inertial range in the around the interval $10 \lesssim k \lesssim 50$.



According to Nazarenko \cite{nazarenko}, 3D numerical simulations of Bose-Einstein condensation (BEC) phenomenon within the nonlinear Schr{\"o}dinger (NLS) equation presents a gradual transition between a KZ power-law energy cascade to a Bose-Einstein distribution with chemical potential $\mu=0$, that is a Planck's distribution. This author cites \cite{nore_97a,nore_97b,kobayashi_05a,kobayashi_05b,kobayashi_07,yepez_09}, where was found evidences of KZ power law spectra with $\gamma \approx 5/3$. These authors argue that the presence of KZ power laws in 3D NLS models is natural, because of the similarity between these models and Euler's hydrodynamics systems observed by Madelung transformations \cite{nazarenko}. 

This kind of structure is very similar that we have found in our simulations. More specifically, in the top component of $P_\rho\left(k\right)$. This fact seems to be very reasonable from a physical perspective. Micha and Tkachev \cite{micha_04} identify the homogeneous component of the inflaton scalar field, that is the energy source in the driven turbulent stage, as a Bose-Einstein condensed. Besides representing waves, the Fourier modes of the inflaton spectral expansion are related to creation and annihilation operators. Naturally, the major part of the created particles should be bosons. Thus, the turbulent process in the preheating of the Universe should have a direct connection with the wave turbulence in the BEC phenomenon. In the next subsection, we show other aspects related to this fact and observed in the power spectra in the time domain.

We noticed that the power spectrum of the energy density, $P_\rho\left(k\right)$, exhibits two components and the division between these components occurs in $k \approx 30$. In the present simulations, we used the grid resolution $N=32$ and the maximum value for $k$ is $k_{max} = 16\sqrt{3} \approx 27.7$. Nazarenko \cite{nazarenko} comments about the consequences of a finite Fourier expansion on the numerical simulations of the type we performed here, and it might be an explanation for the split of the energy power spectrum near $k=k_{max}$. We interpret the second component lying at the bottom of the spectrum as describing a distribution of created particles that tend to thermalization, where the energy distribution is associated to the Planck's law with the same temperature of the reheated Universe, $T_{reh}$.


\subsection{Power spectra in time domain}

%

In addition to the analysis of the power spectra in the spatial domain, we can also obtain insights about the energy flux from the inflaton to the particle production after analyzing the power spectra in time or the frequency domain. We have evaluated tha Fourier expansion of the variance $\sigma_\phi^2$, given by
\begin{equation}
\sigma_\phi^2\left(\tau\right) = \sum_{j=-N/2+1}^{N/2} \widehat{s}_j \expt^{\omega_j \tau},
\end{equation}

\noindent where $\omega_j = 2\pi j/\left[\left(N-1\right)h_e\right]$ and $h_e$ is related to the discretization in the time domain. To obtain the best possible resolution in these spectra, we considered $h_e = h$. The graphics of the power spectra $P_\sigma\left(\omega\right)$ were built from the independent modes of the Fourier expansion of the variance expansions that are in the range  $0 \leqslant j \leqslant N/2$ due to their symmetries.

\begin{figure}[h]
\begin{center}
\includegraphics[scale=0.61]{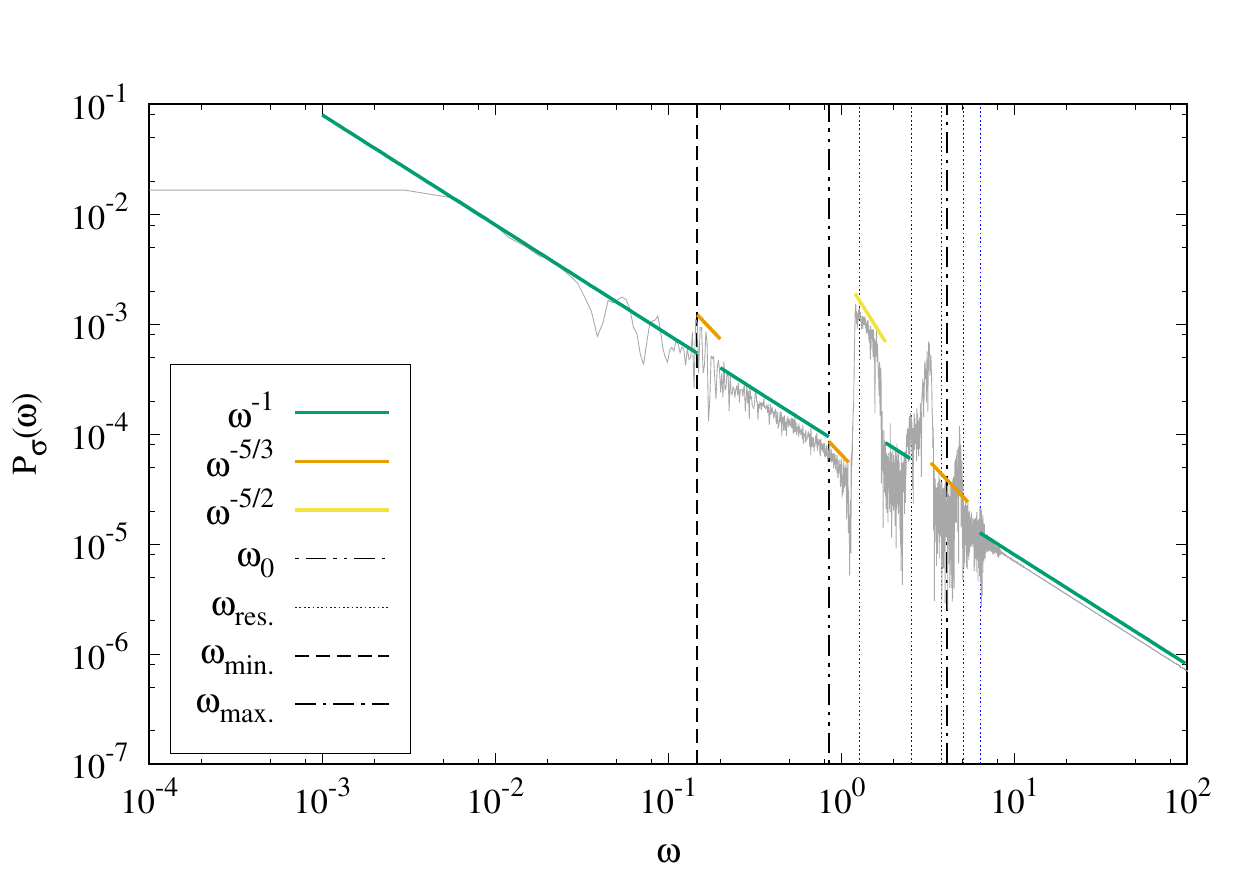}
\includegraphics[scale=0.61]{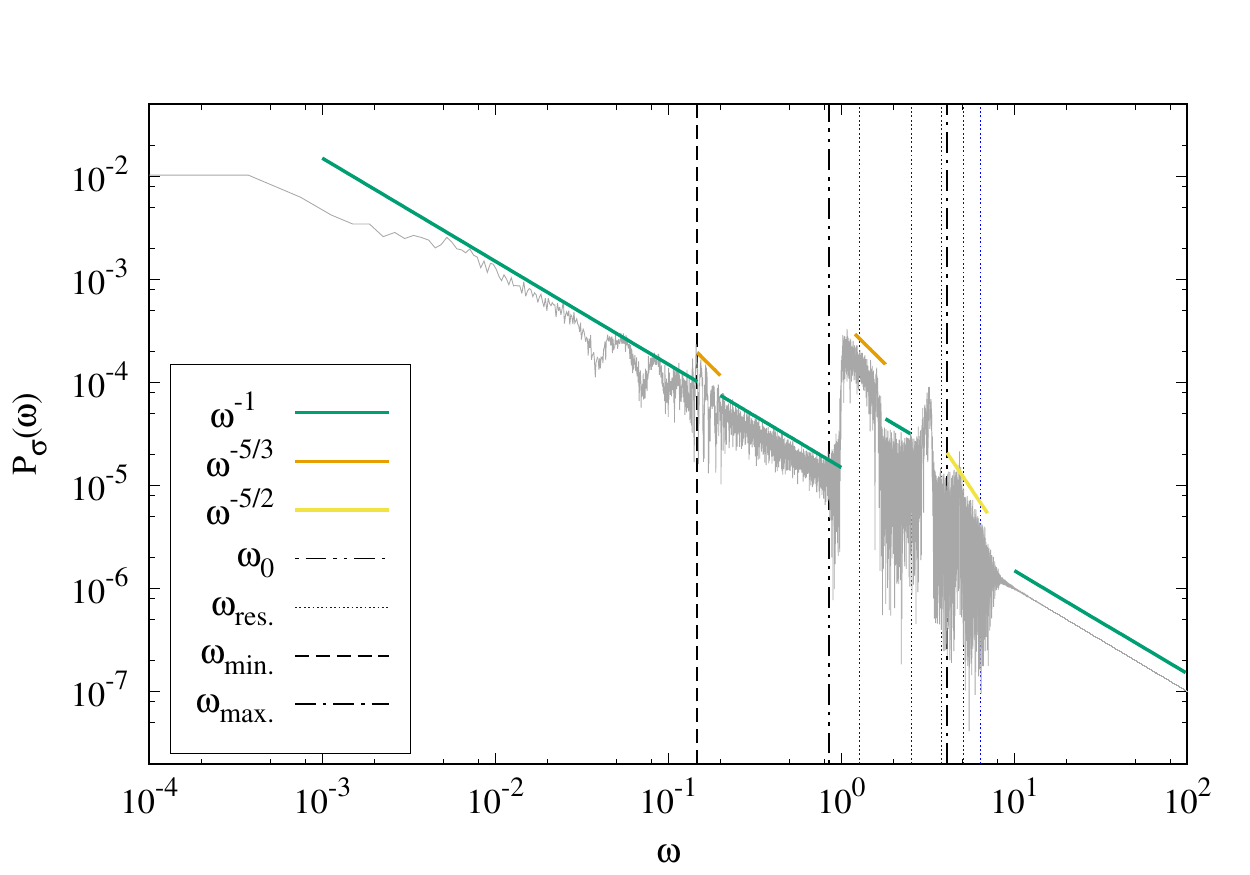}
\end{center}
\caption{\label{vps_0}Power spectra in time domain, $\xi = 0$.}
\end{figure}

\begin{figure}[h]
\begin{center}
\includegraphics[scale=0.61]{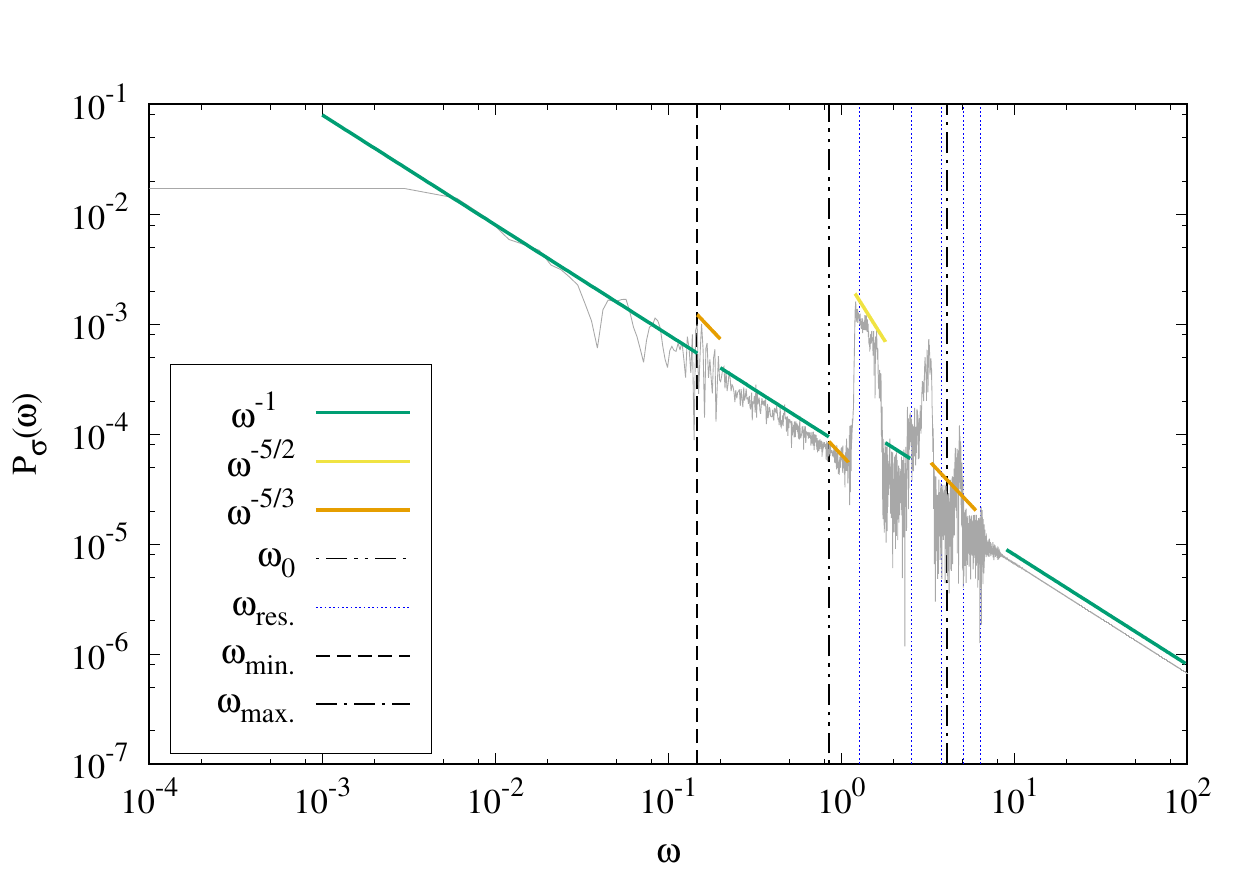}
\includegraphics[scale=0.61]{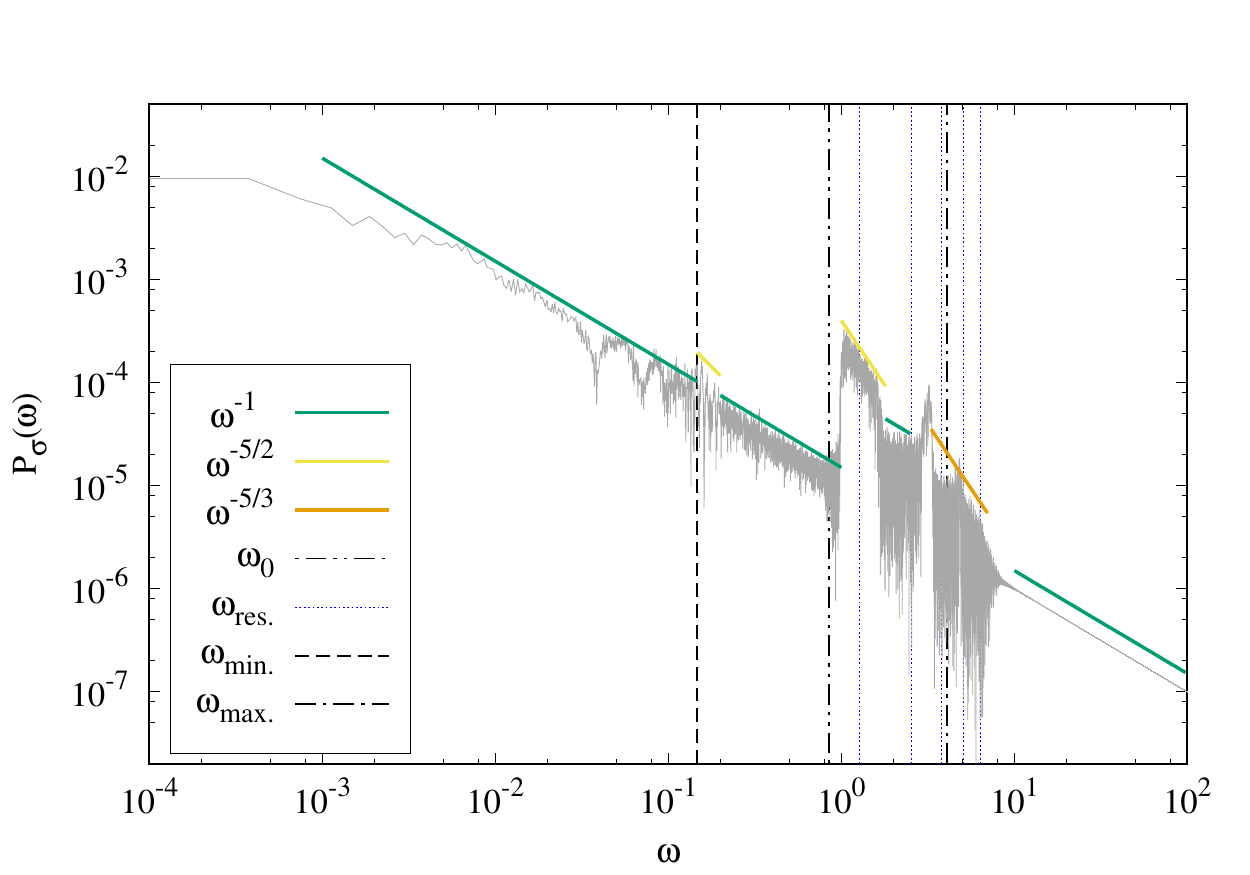}
\end{center}
\caption{\label{vps_01}Power spectra in time domain, $\xi = -0.1$.}
\end{figure}

\begin{figure}[h]
\begin{center}
\includegraphics[scale=0.61]{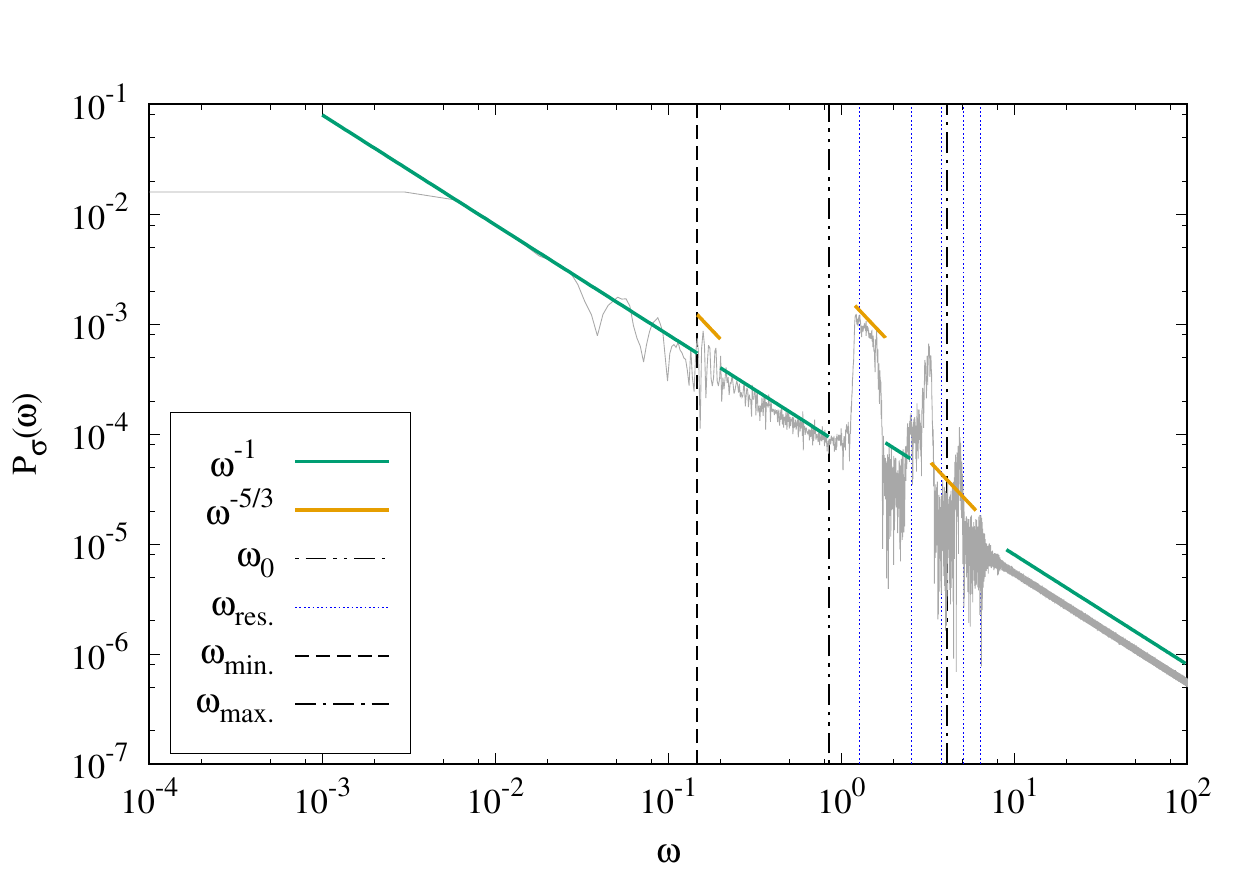}
\includegraphics[scale=0.61]{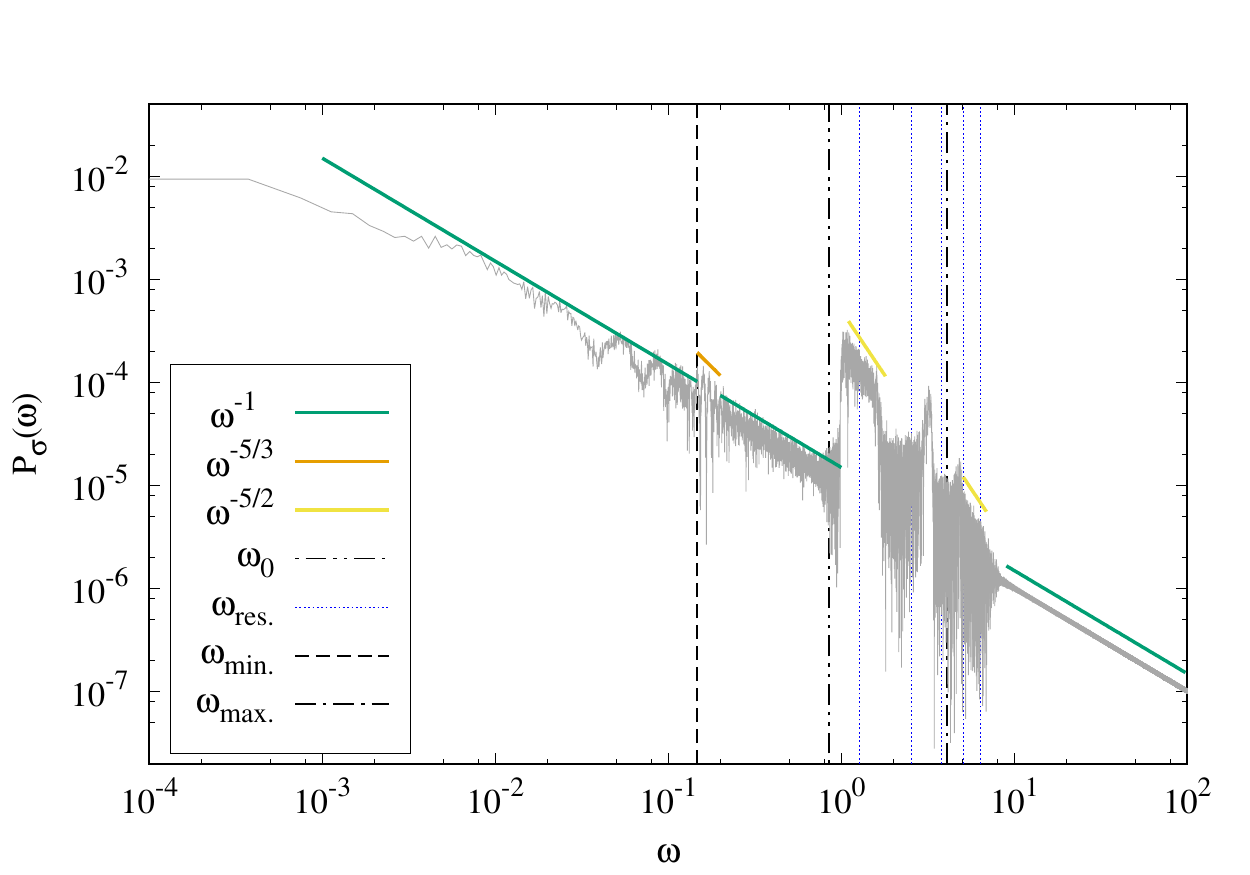}
\end{center}
\caption{\label{vps_1}Power spectra in time domain, $\xi = -1$.}
\end{figure}

The Figs. \ref{vps_0}, \ref{vps_01} e \ref{vps_1} show the power spectra corresponding to the intervals $\Delta \tau = 2^{21}h = 2097.152$ (panels on the left) and $\Delta \tau = 2^{24}h$ (panels on the right), for $\xi=0$, $\xi=-0.1$ e $\xi=-1$, respectively. The feature shared by all power spectra is the presence of the Kolmogorov-Zakharov power law described by
\begin{equation}
P\left(\omega\right) \propto \omega^{-\gamma},
\end{equation}

\noindent with $\gamma$ assuming distinct values such as  $\gamma \approx 1, 5/3, 5/2$ depending on the ranges of $\omega$. The presence of the above power-law is a signature of any chaotic system; however by trying to understand the origin of these exponents can reveal valuable physical information.
In general, in all physical wave systems characterized by a unidimensional parameter (for example, the wave number $k = \left\vert \mathbf{k} \right\vert$) there exists a dispersion relation given by
\begin{equation}
\omega = \lambda_0 k^\alpha,
\end{equation}

\noindent where the physical aspects of the system determine the constants $\lambda_0$ and $\alpha$. This feature allows the determination of the scaling law exponents for the energy, $\gamma_E$, and particles, $\gamma_N$, fluxes by the expressions (cf. \cite{nazarenko})
\begin{eqnarray}
\gamma_E & = & 6 - d - 2\alpha - \dfrac{5-d-3\alpha}{N_w - 1},\label{e-flux} \\
\gamma_N & = & 6 - d - \dfrac{5-d-2\alpha\left(N_w - 2\right)}{N_w - 1}, \label{n-flux}
\end{eqnarray}

\noindent where $d$ is the number of the spatial domain dimensions in the models and $N_w$ is the number of the nonlinear interacting waves. In the preheating case, we have $d=3$ and $\alpha = 1$ \cite{micha_03,micha_04} and the Eqs. \ref{e-flux} and \ref{n-flux} give us
\begin{equation}
\gamma_E = \dfrac{N_w}{N_w - 1} \qquad \textrm{and} \qquad \gamma_N = 1. \label{exp-flux}
\end{equation}

\noindent If $\gamma = \gamma_E$, the power spectra represents an inverse KZ energy cascade; if $\gamma = \gamma_N$, we have a double cascade, where there are a direct energy cascade and an inverse wave cascade. It means there is an energy flux, but the particle number is conserved, and this kind of behavior can drive the system to a BEC phenomenon \cite{nazarenko}.

Nazarenko \cite{nazarenko} mentions the double cascade has physical meaning that are occurring energy transfer in direction of the greater energy levels and a far from equilibrium particle transfer to the lower energy levels. We noticed that the exponent $\gamma = 1$ is independent of the kind of nonlinear interaction for 3D spatial models. In fact, this exponent is present in all power spectra, mainly in the ranges $\omega < \omega_{min}$ and $\omega > \omega_{max}$. This behavior takes place in generalized quantum models of Bose-Einstein gas, where it represents a direct energy cascade that originated in the condensates with lower $\omega$ values in direction to a range described by a thermal cloud for higher $\omega$ values. The description for lower $\omega$ values is given by a classical wave model and for higher $\omega$ values is given by classical particle models. This effect is called ``wave-particle crossover'' \cite{nazarenko}. In particular for these models, the homogeneous mode frequency is $\omega_{0} = 2 \pi/ \Pi_{\phi}$, the minimum and maximum frequencies, respectively $\omega_{min}$ and $\omega_{max}$, of the spectral expasion are $\omega_{min.} = 2 \pi/ L_p$, $\omega_{max.} = 32 \pi \sqrt{3} / L_{p}$ and the resonant frequencies in the linear stage are $\omega_{res.} = c_p 2 \pi \sqrt{74} / L_p$, with $c_p \in \mathbb{N}$.


Another interesting point mentioned by Nazarenko \cite{nazarenko} is the case of a system without dissipation on the lowest energy levels in which the inverse cascade represents a transient state and is characterized by an auto-similar solution. This solution is represented by a power law with $\gamma \approx 2.48 \approx 5/2$, for $d=3$, that is different of the expected value for a KZ inverse cascade ($\gamma = 7/6$ for $d=3$). It is a typical situation for power spectra with finite capacity in magneto-hydrodynamics models. Based on these facts and knowing that $\gamma \approx 5/3$ is a universal exponent in the Kolmogorov's theory, we can conclude that the KZ energy cascades dominate in the interval $\omega_{min} < \omega < \omega_{max}$. The ranges in frequency with $\gamma \approx 5/3$ are related to direct cascades, and those with $\gamma \approx 5/2$ are related to inverse cascades that emerge from the auto-similar evolution cited by Micha and Tkachev in \cite{micha_03}. These cascades are concentrated around the $\omega_{min}$ and $\omega_{max}$, $\omega_0$ and $\omega_{res}$ .


When we compare the power spectra for different $\Delta \tau$, we can see that these cascade patterns are changing and the laws with $\gamma \approx 5/2$ and $\gamma \approx 1$ become more relevant that $\gamma \approx 5/3$. We believe this fact has a direct relation with the stage of turbulent transition as described in \cite{micha_04}. For the lowest $\Delta \tau$ we noted that stronger energy fluxes between the characteristics frequencies of the spectral expansion and major power laws with $\gamma \approx 1$ are out of this range. This aspect should be related to typical energy cascades of the turbulent process in the BEC phenomenon described by Nazarenko in \cite{nazarenko}. For greater $\Delta \tau$, we have observed a tendency to power laws with $\gamma \approx 1$ and $\gamma \approx 5/2$. This fact can indicate that the driven turbulence stage is finishing, since the number of ranges with stationary fluxes is decreasing, and the free turbulence stage is establishing when the BEC effects are dominants, and the Universe will be driven to a thermalized state.


Again, we shall mention the possible influence of the nonminimal coupling on this process. We noted that all power spectra in the time domain, with the same $\Delta \tau$, present a very similar structure for $\xi = 0$ and $\xi=-0.1$. For $\xi = -1$ there is subtle change of the spectrum structure, with the more predominance of power laws with $\gamma \approx 5/3$ in $\omega_{min.} < \omega < \omega_{max.}$. For greater $\Delta \tau$, all power spectra tend to have a similar power law structure, independently of $\xi$ value. 

\section{Energy distribution temperature and Universe thermalization}

Following the same approach that we have done for two spatial dimensions \cite{crespo_14}, we can obtain the temperature of the energy distribution for differents values of $\xi$. In order to estimate this value, we shall restore the physical variables in the exponential argument, where we have $bk = \hbar c k_{phys}/k_{B}T$, $k_{phys}$ is the physical \emph{momentum}, $k_{B}$ é is the Boltzmann's constant and $T$ is the temperature of the distribution. For nonminimal coupling models, we have $k = L_p k_{phys} /\sqrt{\lambda}\phi_0$ and
\begin{equation}
T = \dfrac{\hbar c \sqrt{\lambda}\phi_0}{k_{B}bL_p} = \left\lbrace \begin{array}{l l} 8.31\times10^{11} \mbox{ GeV} ,& \xi = 0 \\ 6.17\times 10^{11} \mbox{ GeV} ,& \xi=-0.1 \\ 2.87 \times 10^{11} \mbox{ GeV} ,& \xi=-1 \\ 9.70 \times 10^{10} \mbox { GeV} ,& \xi = -10 \end{array}\right.
\end{equation}

\noindent We note a decrease of the temperature as the parameter $\xi$ increases in absolute value. For $\xi=0$ the temperature is about $T \sim 10^{12}\, \mathrm{GeV}$ that is smaller than $T \sim 10^{14}\, \mathrm{GeV}$ we obtained for the model in two-dimensions \cite{crespo_14}. Two factors explain this discrepancy: the difference of $L_p$ from $11.12$ (2d model) to $42.70$ in the present 3d model (the parameters $\phi_0$ and $b$ are unchanged) and a forgotten factor $8 \pi$ due to the use of reduced Planck mass. Then, the present value of $T(\xi=0) \sim 10^{12}\, \mathrm{GeV}$ is consistent. 

The obtained temperatures of the energy distribution might agree with the reheating temperature estimates, in short, $T_{\mathrm{reh}}$. The precise value of $T_{\mathrm{reh}}$ depends on the underlying microphysical processes; nevertheless, an upper bound of $T_{\mathrm{reh}} < 10^{15}\, \mathrm{GeV}$ is estimated in Ref. \cite{lozanov}, but distinct upper and lower bounds are possible such as $10^{8}\, \mathrm{GeV} \lesssim T_{\mathrm{reh}} \lesssim 10^{10}\, \mathrm{GeV}$ \cite{allahverdi} or $10^{8}\, \mathrm{GeV} \lesssim T_{\mathrm{reh}} \lesssim 10^{11}\, \mathrm{GeV}$ \cite{butchmiller}. As we have mentioned, the observation rule out the minimally coupled one field models, but left enough room to validate the nonminimally coupled fields with sufficient small $\xi$. In this case, the obtained temperatures $T_{\mathrm{reh}}$ are in agreement with the upper bounds of the reheating temperature. 

\section{Discussion}

We presented the consequences of the nonlinear stages of preheating after inflation considering a single nonminimally coupled scalar field with quartic potential. The recent study of inflationary models \cite{planck13} showed that even with a small coupling parameter $|\xi| > 10^{-3}$ this single field model is not ruled out by the observations. We integrated the field equations starting from the end of inflation and taken into account the backreaction of the inhomogeneous fluctuations of the scalar field. For this aim, we have extended the code of Ref. \cite{crespo_14} a three-dimensional spatial domain modeled by a box of periodic boundary conditions. 

We explored the wave turbulence features through the power spectra of the variance and the energy density of the scalar field either in time and in function of the wave number. As a consequence of the energy transfer from the homogeneous inflaton field to the inhomogeneous fluctuations in the turbulent phase, some intervals of the power spectra are described by power-laws $k^{-\gamma}$ and $\omega^{-\gamma}$, with $k$ and $\omega$ representing the wavenumber and the frequency, respectively. 

There are two other results worth of mentioning. The first is the equation of state arising from the average spatial procedure of the energy-momentum tensor that received the contribution of the inhomogeneous fluctuations of the inflaton field. We found that after the turbulent regime has been established the matter content behaves as a radiation fluid with $w(\tau) \approx 1/3$ for several values of the parameter $\xi$. The second result refers to the estimate of the reheating temperature from the power spectrum of the energy density in the wavenumber identified as the Planck law describing the radiation spectrum of a black body. We estimated the temperature associated with this spectrum the value $T \approx 10^{11}\,\mathrm{GeV}$ for $\xi$ inside the acceptable observational constraints, and is acceptable as a reheating temperature.

We continue the present investigation by studying the aspects of wave turbulence in the nonlinear stages of preheating in a distinct class of two-fields models in the next paper.

\section*{Acknowledgments}
J. A. Crespo acknowledges the financial support of the Brazilian agency Funda\c c\~ao Carlos Chagas Filho de Amparo \`a Pesquisa do Estado do Rio de Janeiro (FAPERJ). H. P. de Oliveira thanks to Conselho Nacional de Desenvolvimento Cient\'ifico e Tecnol\'ogico (CNPq) and Funda\c c\~ao Carlos Chagas Filho de Amparo \`a Pesquisa do Estado do Rio de Janeiro (FAPERJ) (Grant No. E-26/202.998/518 2016 Bolsas de Bancada de Projetos (BBP)).

\section*{References}


\providecommand{\newblock}{}
\begin{thebibliography}{10}
\expandafter\ifx\csname url\endcsname\relax
  \def\url#1{{\tt #1}}\fi
\expandafter\ifx\csname urlprefix\endcsname\relax\def\urlprefix{URL }\fi
\providecommand{\eprint}[2][]{\url{#2}}

\bibitem{henrique13}
Crespo J~A and de~Oliveira H 2014 {\em Journal of Cosmology and Astroparticle
  Physics\/} {\bf 2014} 006

\bibitem{dufaux2006}
Dufaux J~F, Felder G~N, Kofman L, Peloso M and Podolsky D 2006 {\em Journal of
  Cosmology and Astroparticle Physics\/} {\bf 2006} 006

\bibitem{henrique03}
de~Oliveira H~P and Soares I~D 2003 {\em General Relativity and Gravitation\/}
  {\bf 35} 2079--2087

\bibitem{henrique06}
de~Oliveira H~P and Soares I~D 2006 {\em Journal of Cosmology and Astroparticle
  Physics\/} {\bf 2006} 002

\bibitem{micha02}
Micha R and Tkachev I~I 2003 {\em Physical Review Letters\/} {\bf 90} 121301
  (\textit{Preprint} \eprint{http://arxiv.org/abs/hep-ph/0210202})

\bibitem{micha04}
Micha R and Tkachev I~I 2004 {\em Phys. Rev. D\/} {\bf 70}(4) 043538

\bibitem{planck13}
{Planck Collaboration}, {Ade} P~A~R, {Aghanim} N, {Armitage-Caplan} C, {Arnaud}
  M, {Ashdown} M, {Atrio-Barandela} F, {Aumont} J, {Baccigalupi} C, {Banday}
  A~J and et~al 2013 {\em ArXiv e-prints\/}
  \urlprefix\url{http://arxiv.org/abs/1303.5082}

\bibitem{planck2016}
{Planck Collaboration}, {Ade} P~A~R, {Aghanim} N, {Arnaud} M, {Arroja} F,
  {Ashdown} M, {Aumont} J, {Baccigalupi} C, {Ballardini} M, {Banday} A~J and
  et~al 2016 {\em Astronomy and Astrophysics\/} {\bf 594} A20
  (\textit{Preprint} \eprint{1502.02114})

\bibitem{linde13}
Kallosh R and Linde A 2013 {\em Journal of Cosmology and Astroparticle
  Physics\/} {\bf 6} 027 (\textit{Preprint} \eprint{1306.3211})

\bibitem{okada2014}
Okada N, Nefer~{\c S}eno{\u g}uz V and Shafi Q 2014 {\em ArXiv e-prints\/}
  (\textit{Preprint} \eprint{1403.6403})

\bibitem{tsujikawa00}
Tsujikawa S, Maeda K~i and Torii T 2000 {\em Phys. Rev. D\/} {\bf 61}(10)
  103501

\end{thebibliography}


\begin{thebibliography}{99}
	
\bibitem{planck18} Akrami Y et al., 2018 \textit{Planck 2018 results. X. Constraints on inflation}, [arXiv:1807.06211]. 

\bibitem{planck13} Ade P A R  et al., 2014 Astr. and Astrophys. \textbf{571} A22 [arXiv:1303.5082].

\bibitem{basset} Bassett B. A., Tsujikawa S and Wands D, 2006 Rev. Mod. Phys. \textbf{78} 537 [astro-ph/0507632] and references therein.

\bibitem{micha_03} Micha R and Tkachev I I, 2003 Phys. Rev. Lett. \textbf{90} 121301.

\bibitem{micha_04} Micha R and Tkachev I I, 2004 Phys. Rev. D \textbf{70} 043538.

\bibitem{deol_ivano_03} de Oliveira H P and Soares I D, 2003 Gen. Rel. Grav. \textbf{35} 2079. 

\bibitem{dufaux_06} Dufaux J F, Felder G N, Kofman L, Peloso M. and Podolsky D, 2006 JCAP \textbf{7} 06 [hep-ph/0602144].

\bibitem{felder_06} Felder G N and Kofman L, 2007 Phys. Rev. D \textbf{75} 043518.

\bibitem{crespo_14} Crespo J A and  and de Oliveira H. P., (2014) JCAP \textbf{6} 006 [arxiv:1406.1088]. 

\bibitem{kleb_97} Khlebnikov S Y and Tkachev I I, 1996. Phys. Rev. Lett. \textbf{77} 219.

\bibitem{prokopec} Prokopec T and Roos T G, 1997 Phys. Rev D \textbf{55} 3768.  

\bibitem{okada} Okada N, Rehman M U and Shafi Q, 2010 Phys. Rev D \textbf{82} 043502.

\bibitem{kaloshi} Kallosh R and Linde A, 2013 JCAP \textbf{06} 027 [hep-th/1306.3211v2].

\bibitem{tsujikawa} Tsujikawa S, Maeda K and Torii T, 2000 Phys. Rev. D \textbf{61} 103501.

\bibitem{boyd} Boyd J, 2001 \textit{Chebyshev and Fourier spectral methods} (New York: Dover Publications).

\bibitem{fakir} Fakir R and Unruh W G, 1990 Phys. Rev. D \textbf{41} 1783.

\bibitem{latticeeasy} G. N. Felder and I. Tkachev, Comp. Phys. Commun. \textbf{178}, 929 (2008).

\bibitem{defrost} A. V. Frolov, JCAP, \textbf{0811}, 009 (2008).
	
\bibitem{pspectre} R. Easther, H. Finkel and N. Roth, JCAP, \textbf{1010}, 025 (2010).

\bibitem{terry_09} Terry P W and Tangri V, 2009 Phys. of Plasmas \textbf{16} 082305.

\bibitem{nazarenko} Nazarenko S, 2011 \textit{Wave turbulence} (Berlin: Springer)

\bibitem{nore_97a} Nore C, Abid M and Brachet M E, 1997 Phys. Rev. Lett. \textbf{78} 3896.

\bibitem{nore_97b} Nore C, Abid M and Brachet M E, 1997 Phys. of Fluids \textit{9} 2644.
	
\bibitem{kobayashi_05a} Kobayashi M and Tsubota M, 2005 Phys. Rev. Lett. \textbf{94} 065302.

\bibitem{kobayashi_05b} Kobayashi M and Tsubota M, 2005 J. of the Phys. Soc. of Japan \textbf{74} 3248.
	
\bibitem{kobayashi_07} Kobayashi M and Tsubota M, 2007 Phys. Rev. A \textbf{76} 045603.

\bibitem{yepez_09} Yepez J, Vahala G, Vahala L and Soe M, 2009 Phys. Rev. Lett. \textbf{103} 084501.

\bibitem{lozanov} Lozanov K, 2018 \textit{Lectures on Reheating after Inflation}, 2018 MPA Lecture Series in Cosmology  $[https://wwwmpa.mpa-garching.mpg.de/\sim komatsu/lecturenotes/Kaloian\_Lozanov\_on\_Reheating.pdf].$


\bibitem{allahverdi} Allahverdi R, 2000 Phys. Rev. D \textbf{62} 063509.

\bibitem{butchmiller} Buchmuller W, 2012 \textit{Baryogenesis, Dark Matter and the Maximal Temperature of the Early Universe}, Lectures presented at the LII Cracow School of Theoretical Physics, May 2012, Zakopane, Poland [hep-th/1212.3554].


	
\end{thebibliography}

\end{document}